\documentclass[journal]{IEEEtran}
\usepackage[T1]{fontenc}
\usepackage{float}
\usepackage{amsmath}
\usepackage{amsthm}
\usepackage{amssymb}
\usepackage{stackrel}
\usepackage{graphicx}
\usepackage{setspace}
\usepackage{url}
\allowdisplaybreaks

\usepackage{caption}
\usepackage{subfigure}
\usepackage{algorithm}
\usepackage[graphicx]{realboxes}
\usepackage{algorithmic}
\usepackage{verbatim}
\renewcommand{\algorithmicrequire}{\textbf{Input:}}
\renewcommand{\algorithmicensure}{\textbf{Output:}} 
\makeatletter

\floatstyle{ruled}
\newfloat{algorithm}{tbp}{loa}
\providecommand{\algorithmname}{Algorithm}
\floatname{algorithm}{\protect\algorithmname}
\theoremstyle{plain}

\theoremstyle{definition}

\theoremstyle{plain}

\theoremstyle{plain}
\newcommand{\RNum}[1]{\uppercase\expandafter{\romannumeral #1\relax}}
\usepackage[thicklines]{cancel}
\IEEEoverridecommandlockouts
\usepackage{ifpdf}
\usepackage{cite}
\usepackage{graphicx}
\ifCLASSOPTIONcompsoc
\else
\fi
\usepackage{graphicx}
\usepackage{epstopdf}
\usepackage{mathtools}
\usepackage{booktabs} 
\usepackage{multirow}
\usepackage{setspace}
\usepackage{lettrine} 
\usepackage{bm}
\makeatother
\usepackage{babel}

\usepackage{amsmath}

\newtheorem{lemm}{Lemma}

\usepackage{booktabs}
\usepackage{longtable}
\usepackage{amsthm}
\usepackage{enumitem,textcomp,amsfonts,multicol,epsfig,bbding,cases,dsfont,bbm,tabularx,makecell,lscape,chngcntr,mwe,adjustbox}
\usepackage{array}
\usepackage{mdwmath}
\usepackage{mdwtab}
\usepackage{eqparbox}
\usepackage{upgreek}

\usepackage{stfloats}
\usepackage{afterpage}
\usepackage{mathrsfs}
\usepackage{xcolor}

\begin{document}
\captionsetup[figure]{font={small}, name={Fig.}, labelsep=period}
\title{Stacked Intelligent Metasurfaces-Based Electromagnetic Wave Domain Interference-Free Precoding}
    \author{
    Hetong Wang, Yashuai Cao, Tiejun Lv,~\IEEEmembership{Senior Member,~IEEE},
    Jintao Wang,~\IEEEmembership{Fellow,~IEEE},\\
    Ni Wei, Jiancheng An,~\IEEEmembership{Member,~IEEE}, and Chau Yuen,~\IEEEmembership{Fellow,~IEEE}

    \thanks{Manuscript received 4 March 2025; revised 20 November 2025; accepted 14 January 2026. This paper was supported in part by the National Natural Science Foundation of China under No. 62271068, MOE (Ministry of Education, Singapore), under MOE Tier 2 Award number T2EP50124-0032.
    (\emph{corresponding author: Tiejun Lv}.)}
    \thanks{
    Hetong Wang and Tiejun Lv are with the School of Information and Communication Engineering, Beijing University of Posts and Telecommunications (BUPT), Beijing 100876, China (e-mail: \{htwang\_61, lvtiejun\}@bupt.edu.cn)
    }
    \thanks{Y. Cao is with the School of Intelligence Science and Technology, University of Science and Technology Beijing, Beijing 100083, China (e-mail: caoys@ustb.edu.cn).}
    \thanks{
    Jintao Wang is with the Department of Electronic Engineering, Tsinghua University, Beijing 100084, China, also with the State Key Laboratory of Space Network Communications, Beijing 100084, China, and also with the Beijing National Research Center for Information Science and Technology (BNRist), Beijing 100084, China (e-mail: wangjintao@tsinghua.edu.cn).
    }
    \thanks{Ni Wei is with the School of Information Science and Engineering, Fudan University, Shanghai 200433, China (e-mail: weini@fudan.edu.cn).}
    \thanks{Jiancheng An and Chau Yuen are with the School of Electrical and Electronics Engineering, Nanyang Technological University, Singapore 639798 (e-mail: jiancheng\_an@163.com, chau.yuen@ntu.edu.sg).}
}

\maketitle
\begin{abstract}
This paper introduces an interference-free multi-stream transmission architecture leveraging stacked intelligent metasurfaces (SIMs), from a new perspective of interference exploitation.
Unlike traditional interference exploitation precoding (IEP) which relies on computational hardware circuitry, we perform the precoding operations within the analog wave domain provided by SIMs. 
However, the benefits of SIM-enabled IEP are limited by the nonlinear distortion (NLD) caused by power amplifiers.
A hardware-efficient interference-free transmitter architecture is developed to exploit SIM's high and flexible degree of freedom (DoF), where the NLD on modulated symbols can be directly compensated in the wave domain.
Moreover, we design a frame-level SIM configuration scheme and formulate a max-min problem on the safety margin function.
With respect to the optimization of SIM phase shifts, we propose a recursive oblique manifold (ROM) algorithm to tackle the complex coupling among phase shifts across multiple layers.
A flexible DoF-driven antenna selection (AS) scheme is explored in the SIM-enabled IEP system.
Using an ROM-based alternating optimization (ROM-AO) framework, our approach jointly optimizes transmit AS, SIM phase shift design, and power allocation (PA), and develops a greedy safety margin-based AS algorithm.
Simulations show that the proposed SIM-enabled frame-level IEP scheme significantly outperforms benchmarks. Specifically, the strategy with AS and PA can achieve a 20 dB performance gain compared to the case without any strategy under the 12 dB signal-to-noise ratio, which confirms the superiority of the NLD-aware IEP scheme and the effectiveness of the proposed algorithm. 
\end{abstract}


\section{Introduction}\label{Sec.I}
\IEEEPARstart{L}{ow} cost, low power consumption, and high-speed transmission are critical requirements for future mobile communications~\cite{9040264, 9261955}. In 2020, the ITU-R WP5D working group assessed the future directions of sixth-generation (6G) development, emphasizing that the key driving force is to improve spectral efficiency and energy efficiency (EE), and reduce power consumption~\cite{ITU2022}. In pursuit of the 6G vision, reconfigurable intelligent metasurface (RIS) has been viewed as a critical technology due to its low cost and high spectral efficiency~\cite{8910627}. 
RISs have shown great potential to improve spatial division multiplexing capabilities~\cite{11203988}. 
However, such a single-layer metasurface for electromagnetic (EM) phase control has a severe efficiency limitation, hindering the practical application of RIS. To overcome the limitation, stacked intelligent metasurfaces (SIMs)~\cite{10158690} have been proposed to provide more degrees of freedom (DoFs) for manipulating EM of signals~\cite{YU20226}. Joint control of both phase and amplitude of SIMs enables more efficient computation in the analog domain~\cite{10571026}. Modulation, precoding, or other signal filtering can be operated without dedicated hardware. \par

Current studies on RIS and SIM focus on block-level precoding (BLP), which aims to suppress interference between signal streams within channel coherence time blocks (or frames). In practice, it is challenging to eliminate interference, which requires substantial and expensive active hardware. To this end, the symbol-level precoding (SLP) technique~\cite{7103338} has been proposed, which converts interference into constructive power. Compared to anti-interference precoding, the key difference is that SLP designs precoding by simultaneously utilizing both channel and data symbol information. This method eliminates inter-stream interference and only requires a low-complexity interference-free receiver. 
However, the SLP can dramatically increase the transmitter complexity due to frequent optimization on a symbol basis. \par

While SIMs can achieve more complicated precoding operations with programmable wave-based analog computing, 
three key challenges arise in SIM-enabled precoding: 1) Existing beam design of SIM under Gaussian input assumptions inevitably leads to a degradation in communication performance, calling for a finite-alphabet-oriented design; 2) hardware impairments may distort the transmission of SIM-enabled wave domain modulation symbols; and 3) the fixed number of antennas in SIM-assisted wireless systems fails to provide additional DoFs for beam control, and lacks the flexibility to adapt to varying propagation environments and dynamic user distributions. \par

\subsection{Related Work}\label{Sec.I.1}
\subsubsection{Symbol-Level and Frame-Level Precoding}\label{Sec.I.1.1}
Compared to anti-interference BLP, classic SLP harvests constructive power from interference by pushing the received digital symbol away from the decision boundary~\cite{9035662}. The distance between the received digital symbol and the decision boundary is defined as the~\textit{safety margin}~\cite{10171154}. By maximizing the safety margin, the symbol error rate (SER) performance can be improved while reducing transmit power consumption.
To reap the benefits of SLP, some SLP works attempt to deal with intensive optimization of precoding per symbol~\cite{8647428, 9770790, 8465957}. However, these methods are still limited to per-slot precoding framework by only simplifying the derivation complexity. \par

To avoid frequent precoding calculations, Li \emph{et al.}~\cite{9962829} applied a constant precoding matrix within a channel-coherent time frame by optimizing the average SER performance across all symbols. As this approach utilizes inter-stream interference to design the precoding per frame, it is termed interference exploitation precoding (IEP). While frame-level IEP alleviates the complexity of traditional SLP, complicated precoding operations depend on massive expensive hardware elements, e.g., high-precision analog-to-digital converters (ADC) and radio frequency (RF) chains. \par

\subsubsection{Single-Layer Intelligent Metasurface-Assisted Precoding}
The RIS offers a low-cost solution to transceiver hardware via passive signal manipulation. Current RIS-assisted communications are mainly designed based on the assumption of Gaussian signal input. It has been shown that beamforming designs for Gaussian inputs are suboptimal for communication systems with finite alphabet sets~\cite{8642953}. 
Motivated by this, reflection-type metasurface-assisted SLP schemes~\cite{9682517, 9139465, 9219206, 8928065, 9435988} have emerged.
In~\cite{9682517}, the base station (BS) adopted the SLP to minimize the transmit power with the cascaded links of RIS.
In~\cite{9139465, 9219206}, a RIS-assisted multi-user multiple input single output (MU-MISO) communication system was studied, where constructive interference (CI) is achieved by the collaboration between the BS and RIS. \par

Notably, several studies have explored using single-layer RIS as a direct precoder.
Instead of operating as a passive relay, the single-layer RIS with an RF generator was integrated in~\cite{8928065}, to serve as a transmitter and modulate information symbols to achieve SLP functionality. This method only supports single-stream transmission due to its limited DoF and hardware capabilities, and inherits the requirement for slot-by-slot switching of the beam pattern.
The authors of~\cite{9435988} extended this method to the mapping from multi-user symbols to the reflection phase shifts using SLP. 
Nevertheless, this system functions as an RF signal generator and lacks capability of signal reception or sophisticated signal processing. \par

Dynamic metasurface antennas (DMAs) have been proposed as a type of reconfigurable antenna~\cite{10584442} for precise beamforming in hybrid precoding by replacing analog phase shifts~\cite{11202489} and jointly optimizing DMA weights with digital precoding vectors.
While DMAs can partially reduce the number of required RF chains and enable simpler and more efficient system designs through their direct, planar, and energy-efficient characteristics~\cite{11202489}, they still rely on baseband digital precoding that incurs high computational complexity and cannot alleviate the need for high-precision digital-to-analog converters (DACs) and other expensive hardware components. \par

\subsubsection{Stacked Intelligent Metasurface-Assisted Precoding}
The advances in few-layer metasurfaces~\cite{201501506} and diffractive neural network~\cite{Liu2022} have inspired efforts in the SIM~\cite{10515204, an2023stacked}, because of its unprecedented capabilities for light-speed calculations with low power consumption. The application of SIM-enabled wave-domain analog computing has recently been explored across various systems.
The multi-stream holographic multiple input multiple output (HMIMO) system was proposed in~\cite{10534211}. It employs SIMs as the precoder and combiner at the transmitter and receiver, respectively. The achievable rate of the HMIMO system was maximized by jointly optimizing phase shifts of the SIMs and the covariance matrix of the transmitted signal.
In~\cite{10543143}, the achievable downlink sum rate was maximized for a SIM-assisted MU-MISO system. This was achieved by alternately optimizing the SIM's phase shifts and the BS's power allocation (PA), based on statistical channel state information (CSI).
The study in~\cite{10683447} explored a SIM-enabled wave-based beamfocusing framework for near-field MU-MISO downlinks. The work minimized the multi-user interference (MUI) by optimizing the diffraction behavior of meta-atoms with reconfigurable amplitude and phase, exploiting the additional angular DoFs inherent in spherical wavefronts.
The SIM composed of phase-controlled and amplitude-controlled layers was investigated in~\cite{10824842} to maximize the sum-rate capacity of a downlink MU-MISO system. The transmit PA at the BS and the amplitude-phase responses of the SIM were jointly optimized to suppress interference among user streams. \par

However, most current SIM-related studies mainly focused on anti-interference precoding design, without considering the practical finite alphabet inputs.
Moreover, derivations related to SIM typically lead to algorithms and conclusions based on ideal hardware assumptions~\cite{10379500}, which neglects the impact of nonlinear distortion (NLD) from power amplifiers.
More recently, the impact of the hardware impairments on the sum-rate has been investigated in the SIM-aided cell-free system~\cite{10535263, 10918608}. While the performance degradation caused by hardware impairments cannot be ignored, the SIM coefficients are still optimized based on Gaussian inputs. Furthermore, power amplifier nonlinearities are modeled as interference, rather than explicitly characterizing the phase and amplitude distortion effects.
Although pre-distortion processing~\cite{bhargava2023pre, 10258313} in traditional multiple input multiple output (MIMO) systems can mitigate the NLD effect, it significantly increases the power consumption and hardware cost of the BS. 

\subsection{Contributions and Organization}\label{Sec.I.2}
In this paper, we present a new SIM-enabled IEP architecture in MU-MISO networks, where signal processing and pre-distortion are implemented in the EM domain.
The contributions of the paper are summarized as follows.
\begin{itemize}
    \item We put forward a SIM-enabled IEP architecture for MU-MISO systems. Through multi-layer transmissive metasurfaces, analog wave-domain precoding is implemented. Thanks to SIM-based acceleration, IEP is expected to operate at much higher speeds than electronic implementations. Rather than suppressing MUI, this novel precoding scheme leverages interference to enhance signal detection.
    \item Unlike existing SIM studies, e.g.,~\cite{liu2024stacked}, we consider non-negligible hardware impairments caused by nonlinear power amplifiers.
    Modeling the NLD as the memoryless nonlinear Saleh model, we design NLD-aware IEP optimization to compensate for NLD on modulated signals in the EM wave, with no need for pre-distortion processing modules at the BS. 
    \item To avoid symbol-by-symbol precoding, we formulate a SIM-assisted EM domain-based frame-level IEP problem. Optimizing the SIM phase shifts is challenging due to the complex coupling of phase shifts among multiple layers. We develop a recursive oblique manifold (ROM) algorithm to obtain the optimal SIM phase shifts that maximize the minimum safety margin.
    \item To achieve flexible DoF, we study the antenna selection (AS) problem under the SIM-enabled IEP architecture. For the target problem, a greedy safety margin-based AS algorithm is proposed. We develop an ROM-based alternating optimization (ROM-AO) framework to address
    the joint transmit AS, SIM phase shift design, and PA problem. Simulation shows that the strategy with AS and PA achieves a 20 dB performance gain compared to the case without any strategy, at a signal-to-noise ratio (SNR) of 12 dB.
\end{itemize}

\begin{figure}[t]
	\centering
    \includegraphics[width=3.5in]{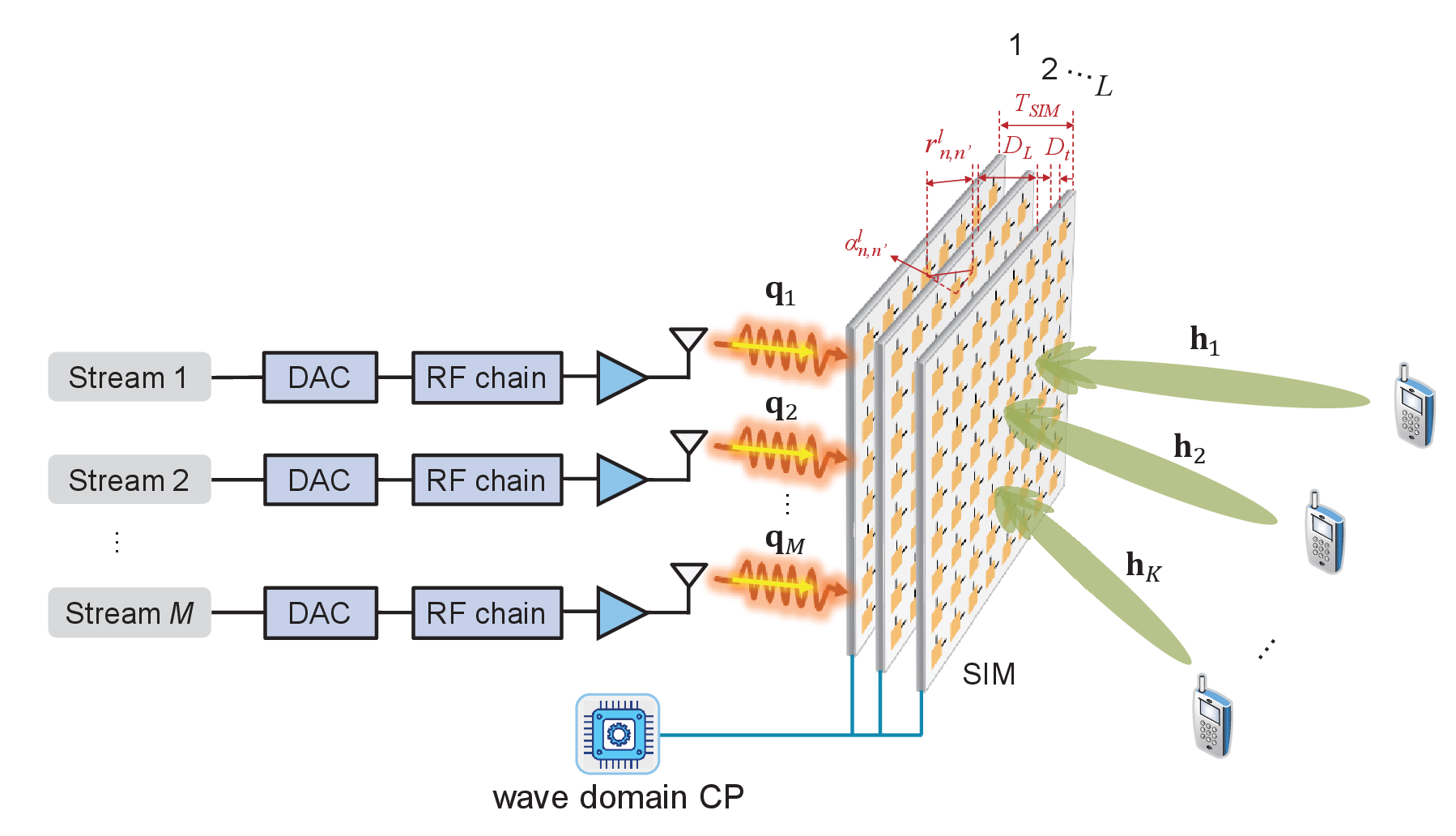}
	\caption{SIM-empowered integrated modulation and precoding framework in the EM wave domain.}
	\label{fig:1}
\end{figure}

The remainder of this paper is organized as follows. Section~\ref{Sec.II} describes the SIM-enabled MU-MISO system and formulates the optimization problem for SIM-enabled EM wave-based IEP. The ROM algorithm for optimizing SIM phase shifts is described in Section~\ref{Sec.III}. The solutions to the AS and PA problems are presented in Section~\ref{Sec.IV}. Section~\ref{Sec.V} demonstrates the performance of the proposed SIM-enabled EM-based IEP scheme. This paper is summarized in Section~\ref{Sec.VI}.
 
\emph{Notations}: various lower- and upper-case boldface letters indicate column-vectors and matrices, respectively; $(\cdot)^{\mathrm{T}}$ and $(\cdot)^{\mathrm{H}}$ denote the transpose and conjugate transpose, respectively; $\angle\{\cdot\}$ returns the angle of a complex value; $\Vert \cdot \Vert_F$ denotes the Frobenius norm (F-norm) of a matrix; $\Re\{\cdot\}$ and $\Im\{\cdot\}$ indicate the real and imaginary parts of complex values, respectively; The superscript $(\cdot)^{t}$ indicates the $t$-th iteration step; $\mathrm{Diag}\{\cdot\}$ represents the diagonal matrix with diagonal entries taken from the elements of the input matrix, and $\mathrm{diag}\{\cdot\}$ represents the extraction of diagonal elements from the matrix into a column vector; $\mathrm{Proj}\{\cdot\}$ denotes the projection operator.

\section{System Model and Problem Formulation}\label{Sec.II}
\subsection{Transmission System}\label{Sec.II.1}
This paper proposes a hardware-efficient MU-MISO downlink transmission system leveraging interference exploitation.
The principle is to integrate a SIM having high reconfigurability with a BS as a transmitter system. Joint control of phase and amplitude responses can be achieved. 
As illustrated in Fig. \ref{fig:1}, the BS is equipped with $M$ antennas, denoted by $\mathcal{M}_0 = \{ 1, \cdots, M\}$, covered by a SIM radome. The SIM is composed of $L$-layer transmissive-type metasurfaces denoted by $\mathcal{L}=\{1,\cdots,L\}$, which are generally embedded into a transparent substrate by well-established top-down nano fabrication techniques~\cite{YU20226}.
The metasurface of each layer contains $N$ meta-atoms denoted by $\mathcal{N}=\{1,\cdots,N\}$, and each tune independently. Each antenna first generates the signal via the RF source. Afterwards, the signals are directed to the first layer of the SIM, and then transmitted through the remaining layers until they penetrate through the $L$-th layer, on which beams are formed and pointed to desired users.

\subsubsection{NLD model for modulated symbols}\label{Sec.II.1.1}
The BS selects $K$ out of the $M$ antennas to serve $K$ users, denoted by $\mathcal{K}=\{1,2,\cdots,K\}$.
The modulated digital symbol vector in the $\mu$-th time slot within a frame $\tilde{\mathbf{s}}_{\mu} = \left[\tilde{s}_{\mu, 1},\tilde{s}_{\mu, 2},\cdots,\tilde{s}_{\mu, K}\right]^{\mathrm{T}}$ includes $K$ independent and identically distributed (i.i.d.) random symbols with zero mean and unit variance. 

In practice, the power amplifiers yield the NLD effect and affect both the amplitude and phase of the modulated signals~\cite{10769490}. The widely-used memoryless nonlinear Saleh model~\cite{10810111,10694880} can describe the amplitude distortion $A(\tilde{s}_{\mu, k})$ and phase distortion $\phi(\tilde{s}_{\mu, k})$ for the modulated symbol $\tilde{s}_{\mu, k}$:
\begin{align}
    A(\tilde{s}_{\mu, k}) &= \alpha_a\vert\tilde{s}_{\mu, k}\vert/\left( 1 + \beta_a\vert\tilde{s}_{\mu, k}\vert^2\right) ;  \label{eq:II.1.1.1} \\
    \phi(\tilde{s}_{\mu, k}) &= \alpha_{\phi}\vert\tilde{s}_{\mu, k}\vert^2 / \left( 1 + \beta_{\phi}\vert\tilde{s}_{\mu, k}\vert^2\right),\label{eq:II.1.1.2}
\end{align}
where $\alpha_a$ and $\beta_a$ are the gain and compression factors of the modulated symbol amplitude, respectively; $\alpha_{\phi}$ and $\beta_{\phi}$ are the gain and compression factors of the modulated symbol phase, respectively. \par

Based on the nonlinear distorted Saleh model~\eqref{eq:II.1.1.1} and~\eqref{eq:II.1.1.2}, the nonlinear distorted symbol $s_{\mu, k}$ can be expressed as
\begin{align}
    s_{\mu, k} = A(\tilde{s}_{\mu, k})e^{j(\angle \tilde{s}_{\mu, k} + \phi(\tilde{s}_{\mu, k}))},\label{eq:II.1.1.3}
\end{align}
where $\angle \tilde{s}_{\mu, k}$ is the argument of $\tilde{s}_{\mu, k}$. The transmitted symbol vector after the NLD is $\mathbf{s}_{\mu}=\left[s_{\mu, 1},s_{\mu, 2},\cdots,s_{\mu, K}\right]^{\mathrm{T}}$. \par

\subsubsection{Channel model for SIM-assisted MU-MISO downlink communication system}\label{sec:II.1.2}
The channel between the $k$-th user and the $L$-th SIM layer is denoted by $\mathbf{h}_k \in \mathbb{C}^{N \times 1}$. The channel between the $l$-th layer and $(l-1)$-th layer of the SIM is denoted by $\mathbf{Q}_l$.
As a result, the equivalent channel between the first layer and the last layer of the SIM can be expressed by
\begin{align}
\mathbf{G} = \mathbf{\Phi}_L \mathbf{Q}_L \mathbf{\Phi}_{L-1} \mathbf{Q}_{L-1} \cdots \mathbf{\Phi}_2 \mathbf{Q}_2 \mathbf{\Phi}_1 \in \mathbb{C}^{N \times N},\label{eq:II.1.2.1}
\end{align}
where $\mathbf{\Phi}_l=\mathrm{Diag} \left\{ \boldsymbol{\theta}_{l} \right\}$ is the phase shift matrix of the $l$-th SIM layer with $\boldsymbol{\theta}_{l} = \left[\theta_{l,1}, \theta_{l,2}, \cdots, \theta_{l,N} \right]^{\mathrm{T}}$ representing the phase shift vector, and $\theta_{l,n} = \mathrm{e}^{\mathrm{j}\phi_{l, n}}$ with $\phi_{l, n} \in [0, 2\pi)$ denoting the corresponding argument. $\mathrm{Diag}\{\cdot\}$ represents the diagonal matrix with diagonal entries taken from the elements of the input matrix. The radiation patterns of the SIM are switched via the wave-domain control processor (CP), as shown in Fig.~\ref{fig:1}.

With the channel model \eqref{eq:II.1.2.1}, the received signal at the $k$-th user in the $\mu$-th time slot within a frame is given by
\begin{align}
y_{\mu, k} = \mathbf{h}_k^{\mathrm{H}} \mathbf{G} \mathbf{Q}_1 \mathbf{s}_{\mu},
\label{eq:II.1.2.2}
\end{align}
where $\mathbf{Q}_1 \in \mathbb{C}^{N \times K}$ denotes the channel between the first SIM layer and the transmit antennas.

\subsection{Scattering Model of Stacked Intelligent Surfaces}\label{sec:II.2}
To analyze channel characteristics in the SIM-empowered interference exploitation system, we first introduce the channel model by applying the Rayleigh-Sommerfeld diffraction theory of EM waves~\cite{wang2016broadband}.
As shown in Fig.~\ref{fig:1}, 
the thickness of the SIM and the spacing between each metasurface are defined as $T_{\mathrm{SIM}}$ and $D_L=T_{\mathrm{SIM}}/L$, respectively. Each layer arranges $N_x$ meta-atoms per row and $N_y$ meta-atoms per column.

Notably, the SIM is realized by metasurfaces having a spatially continuous EM aperture and provides a promising solution for approaching the ultimate capacity limit of wireless channels.
For each layer, the element spacing between adjacent meta-atoms is set to be half-wavelength, i.e., $\Delta \le 0.5 \lambda$. Hence, the size of each layer is given by $N_x\Delta \times N_y \Delta$.
The thickness of each layer, $D_t$, is generally much smaller relative to the separation distance between layers and the element spacing~\cite{akram2020bi}. Hence, the thickness ($D_t \approx 0.045 \lambda$) of the single-layer metasurface~\cite{kim2024screen} can be ignored in this paper~\cite{10423853}. \par

The channel coefficient between the $n$-th meta-atom on the $(l-1)$-th layer and $n'$-th meta-atom on the $l$-th layer, can be modeled by~\cite{wang2016broadband}
\begin{align}
q_{n,n'}^l = \frac{\Delta^2 \cos\alpha_{n,n'}^l }{r_{n,n'}^l}
\left( \frac{1}{2\pi r_{n,n'}^l} - \mathrm{j}\frac{1}{\lambda} \right) \mathrm{e}^{ \mathrm{j}  \kappa r_{n,n'}^l },
\label{eq:II.2.1}
\end{align}
where $\Delta^2$ is the area of each meta-atom, $\alpha_{n,n'}^l$ is the angle between the propagation direction and the normal direction of the $(l-1)$-th layer, $r_{n,n'}^l$ is the propagation distance between the meta-atoms of adjacent layers, and $\kappa=\frac{2\pi}{\lambda}$ is the wavenumber.
Based on \eqref{eq:II.2.1}, the channel $\mathbf{Q}_{1},\mathbf{Q}_{2},\cdots,\mathbf{Q}_{L}$ in~\eqref{eq:II.1.2.1} can be constructed. \par

The channel between the SIM and the $k$-th user is modeled as the correlated Rayleigh fading model under a three-dimensional (3D) set-up~\cite{10130641}. We can write $\mathbf{h}_k$ as
\begin{align}
\mathbf{h}_k \sim \mathcal{CN}\left(\mathbf{0}, \rho_k \mathbf{R} \right),
\label{eq:II.2.2}
\end{align}
where $\rho_k$ is the free space path loss, and the covariance matrix $\mathbf{R}$ is given by
\begin{align}
\left[\mathbf{R} \right]_{n,n'} = \mathrm{sinc} \left( \frac{2 r_{n,n'}^{L}}{\lambda} \right)=\frac{\sin(\kappa r_{n,n'}^{L})}{\kappa r_{n,n'}^{L}}.
\label{eq:II.2.3}
\end{align}
We use $\mathbf{H}=[\mathbf{h}_1, \mathbf{h}_2, \cdots, \mathbf{h}_K] \in \mathbb{C}^{N \times K}$ to collect the channels for all users. Then, the received signal model is given by
\begin{align}
\mathbf{y}_{\mu} = \mathbf{H}^{\mathrm{H}} \mathbf{G} \mathbf{Q}_1 \mathbf{s}_{\mu},
\label{eq:II.2.4}
\end{align}
where $\mathbf{y}_{\mu}=[y_{\mu, 1}, y_{\mu, 2},\cdots, y_{\mu, K}]^{\mathsf{T}}$ collects the received signals for all users in the $\mu$-th time slot within a frame.

\subsection{Transmission Strategy and Problem Formulation}\label{sec:II.3}
\begin{figure}[t]
	\centering{}\includegraphics[width=3.5in]{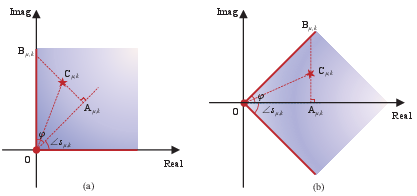}
	\caption{IEP design for QPSK signals: (a) CI region and safety margin for a first-quadrant QPSK point; (b) diagram from Fig. 2(a) after a clockwise rotation by $\angle s_{\mu,k}$.}
	\label{fig:2}
\end{figure}

The high overhead of channel estimation and control signaling makes frequent switching of the SIM phase shifts impractical. Therefore, optimizing the SIM phase shifts symbol-by-symbol to achieve IEP technology would cause unbearable costs. To this end, we design a novel SIM-assisted frame-level IEP strategy. Specifically, we adopt the standard block-fading channel assumption~\cite{9962829,7174558}. The channel remains over a channel coherence time, and experiences independent Rayleigh fading between the SIM and users between frames. Each frame is divided into $U$ consecutive and non-overlapping time units, referred to as slots. Within each slot, $K$ symbols are transmitted simultaneously, one to each of the $K$ users. Accordingly, the SIM phase shifts are configured only once per frame, aligning with the frame-level channel coherence time and reducing the control overhead compared with symbol-level optimization. \par

The IEP technology can convert the MUI signals into constructive power to push digital symbols away from the decision boundaries. Here, the quadrature-phase-shift-keying (QPSK) modulated signals are taken as an example. The CI region of the digital symbol $\left(1/\sqrt{2},\mathrm{j}/\sqrt{2}\right)$ for the $k$-th user in the $\mu$-th time slot within a frame is illustrated in Fig.~\ref{fig:2}. As shown in Fig.~\ref{fig:2}(a), the light purple shadow indicates the CI region, point $\mathrm{C}_{\mu, k}$ represents the received noise-free signal $y_{\mu, k}$, point $\mathrm{A}_{\mu, k}$ represents the projection of point $\mathrm{C}_{\mu, k}$ in the direction of desired symbol $s_{\mu, k}$ with argument $\angle s_{\mu, k}$, point $\mathrm{B}_{\mu, k}$ represents the intersection of $\overrightarrow{\mathrm{AC}_{\mu, k}}$ and the decision boundary, and $\varphi$ represents half of the CI region's angle range, i.e., $\varphi=\mathrm{\pi}/4$. It is observed that when $\vert\overrightarrow{\mathrm{BC}_{\mu, k}}\vert=\vert\overrightarrow{\mathrm{AB}_{\mu, k}}\vert-\vert\overrightarrow{\mathrm{AC}_{\mu, k}}\vert\geq 0$, the received symbol $y_{\mu, k}$ can fall within the CI region, namely, it can be correctly demodulated. Consequently, $\vert\overrightarrow{\mathrm{BC}_{\mu, k}}\vert$ can be utilized to evaluate the performance of IEP, denoted as the safety margin. \par

For simplicity, Fig.~\ref{fig:2}(a) is rotated clockwise by $\angle s_{\mu, k}$ degrees to obtain Fig.~\ref{fig:2}(b). In this case, the vector $\overrightarrow{\mathrm{OA}_{\mu, k}}$, which represents the direction of the desired symbol $s_{\mu, k}$, coincides with the real axis, bisecting the CI region into two symmetric halves. Accordingly, the vector $\overrightarrow{\mathrm{OC}_{\mu, k}}$ for the received noise-free signal $y_{\mu, k}$ is transformed as $\overrightarrow{\mathrm{OC}_{\mu, k}} = y_{\mu, k}e^{-j\angle s_{\mu, k}}$, according to the clockwise rotation $\angle s_{\mu, k}$. To derive the safety margin $\vert\overrightarrow{\mathrm{BC}_{\mu, k}}\vert$, we need to calculate the vector magnitudes $\vert\overrightarrow{\mathrm{AB}_{\mu, k}}\vert$ and $\vert\overrightarrow{\mathrm{AC}_{\mu, k}}\vert$. The vector magnitude $\vert\overrightarrow{\mathrm{AB}_{\mu, k}}\vert$ can be represented as
\begin{equation}
    \vert\overrightarrow{\mathrm{AB}_{\mu, k}}\vert = \Re\left\{y_{\mu, k}e^{-j\angle s_{\mu, k}}\right\}\tan\varphi, \label{eq:II.3.re.1}
\end{equation}
which first obtains the projection of $\overrightarrow{\mathrm{OC}_{\mu, k}}$ onto the real axis, i.e., $\vert\overrightarrow{\mathrm{OA}_{\mu, k}}\vert$.
This projected length is then multiplied by $\tan\varphi$ by considering the right triangle $\bigtriangleup \mathrm{OA_{\mu, k}B_{\mu, k}}$, thereby obtaining $\vert\overrightarrow{\mathrm{AB}_{\mu, k}}\vert$.
We further compute the vector magnitude $\vert\overrightarrow{\mathrm{AC}_{\mu, k}}\vert$ via the projection of $\overrightarrow{\mathrm{OC}_{\mu, k}}$ onto the imaginary axis, which corresponds to
\begin{equation}
    \vert\overrightarrow{\mathrm{AC}_{\mu, k}}\vert = \vert\Im\left\{y_{\mu, k}e^{-j\angle s_{\mu, k}}\right\}\vert. \label{eq:II.3.re.2}
\end{equation}
It is noteworthy that the absolute value operation in the projection onto the imaginary axis guarantees the equivalence of the calculation, regardless of whether the received noise-free signal $y_{\mu, k}$, i.e., point $\mathrm{C}_{\mu, k}$, lies above or below the real axis. Finally, the safety margin $\vert\overrightarrow{\mathrm{BC}_{\mu, k}}\vert$ is calculated as the difference between $\vert\overrightarrow{\mathrm{AC}_{\mu, k}}\vert$ and $\vert\overrightarrow{\mathrm{AB}_{\mu, k}}\vert$ according to~\eqref{eq:II.3.re.1} and~\eqref{eq:II.3.re.2}, which can be expressed as
\begin{equation}
    \vert\overrightarrow{\mathrm{BC}_{\mu, k}}\vert=\Re\left\{y_{\mu, k}e^{-j\angle s_{\mu, k}}\right\}\tan\varphi-\vert\Im\left\{y_{\mu, k}e^{-j\angle s_{\mu, k}}\right\}\vert.\label{eq:II.3.1}
\end{equation}

Since the safety margin $\vert\overrightarrow{\mathrm{BC}_{\mu, k}}\vert$ in~\eqref{eq:II.3.1} directly influences symbol detection performance, it serves as an alternative metric of SER.
Our objective is to maximize the minimum the safety margin, i.e., $\vert\overrightarrow{\mathrm{BC}_{\mu, k}}\vert$ of users in $U$ time slots within a coherent frame, by optimizing the SIM phase shifts $\{\mathbf{\Phi}_l\}_{l=1}^L$. The CSI for all channels can be accurately obtained using the existing channel estimation techniques of SIM~\cite{10445164, 7458188}. The resulting optimization problem is formulated as:
\begin{subequations}
\label{eq:II.3.2}
    \begin{align}
        (\mathrm{P}1):&\max\limits_{\{\mathbf{\Phi}_l\}_{l=1}^L}\ \min\limits_{\mu,k}\ \vert\overrightarrow{\mathrm{BC}_{\mu,k}}\vert \label{eq:II.3.2a}\\
        &\mathrm{s.t.}\ \vert\theta_{l,n}\vert = 1,\forall n\in\mathcal{N},\forall l\in\mathcal{L}, \label{eq:II.3.2b}       
    \end{align}
\end{subequations}
where $\vert\overrightarrow{\mathrm{BC}_{\mu, k}}\vert$ is expressed by~\eqref{eq:II.3.1}.
Due to the discontinuous and non-differentiability objective function, and the constant modulus constraint~\eqref{eq:II.3.2b}, problem ($\mathrm{P}1$) is non-convex. 
While the constant modulus constraint~\eqref{eq:II.3.2b} in single-layer scenarios can be efficiently handled by Riemannian manifold optimization methods~\cite{boumal2014manopt}, directly applying these methods to problem ($\mathrm{P}1$) becomes intractable. This is due to the intricate objective function~\eqref{eq:II.3.2a} and complex coupling among phase shifts across multiple layers.

\section{SIM Phase Shift Optimization} \label{Sec.III}
This section develops the ROM algorithm to optimize SIM phase shifts $\{\mathbf{\Phi}_l\}_{l=1}^L$. The algorithm addresses the core challenges of problem $(\mathrm{P}1)$, which include the coupling of variables across layers in~\eqref{eq:II.3.2b} and the non-convex, non-differentiable objective function~\eqref{eq:II.3.2a}. Our approach follows a structured procedure with two stages. First, problem $(\mathrm{P}1)$ is decoupled into a sequence of single-layer subproblems, thereby handling the coupled constraints. Subsequently, each subproblem transforms the non-differentiability and discontinuity of the objective function~\eqref{eq:II.3.2a} into a form suitable for Riemannian manifold optimization that can be solved in Riemannian space as an unconstrained problem. \par

\subsection{Problem Transformation and Decomposition}\label{Sec.III.1}
The primary challenge in solving problem $(\mathrm{P}1)$ stems from the coupled phase shifts across all $L$ SIM layers. We adopt a recursive strategy that optimizes one layer per step while keeping the others fixed. This approach effectively decouples $(\mathrm{P}1)$ into a sequence of manageable subproblems, where the variable coupling imposed by constraint~\eqref{eq:II.3.2b} is isolated and resolved layer-by-layer. \par

\subsubsection{Problem Decoupling via Layer-Wise Reformulation}
Taking the phase shifts in the $l$-th layer (i.e., $\mathbf{\Phi}_l$) for example, the received signal $y_{\mu, k}$ in~\eqref{eq:II.1.2.2} is expanded as
\begin{align}
y_{\mu, k} = \underbrace{\mathbf{h}_k^{\mathrm{H}} 
\mathbf{\Phi}_L \mathbf{Q}_L \cdots \mathbf{Q}_{l+1}}_{\text{equivalent left channel}} \mathbf{\Phi}_l \underbrace{\mathbf{Q}_{l} \cdots \mathbf{\Phi}_1
\mathbf{Q}_1 \mathbf{s}_{\mu}}_{\text{equivalent right channel}}.
\label{eq:III.1.1}
\end{align}
For ease of illustration, we define 
\begin{equation}
    \tilde{\mathbf{g}}_{1,k,l} \triangleq
    \begin{cases}
    \mathbf{h}_k, & l=1 \\    \mathbf{Q}_{l+1}^{\mathrm{H}}\mathbf{\Phi}_{l+1}^{\mathrm{H}}\cdots \mathbf{Q}_L^{\mathrm{H}} \mathbf{\Phi}_L^{\mathrm{H}}\mathbf{h}_k,
    & l=2,\cdots,L,
    \end{cases}
    \label{eq:III.1.2}
\end{equation}
\begin{equation}
    \tilde{\mathbf{g}}_{2,\mu,k,l} \triangleq 
    \begin{cases}
    \mathbf{Q}_{l} \cdots \mathbf{\Phi}_1
    \mathbf{Q}_1 \mathbf{s}_{\mu},& l=1,\cdots,L-1 \\
    \mathbf{Q}_1 \mathbf{s}_{\mu},& l = L,
    \end{cases}
\label{eq:III.1.3}
\end{equation}
where $\tilde{\mathbf{g}}_{1,k,l}$ denotes the column vector form of the equivalent left channel in~\eqref{eq:III.1.1} and $\tilde{\mathbf{g}}_{2,\mu,k,l}$ denotes the equivalent right channel in~\eqref{eq:III.1.1}.
By using~\eqref{eq:III.1.2} and~\eqref{eq:III.1.3}, Eqn.~\eqref{eq:III.1.1} can be rewritten as
\begin{align}
y_{\mu, k} = \tilde{\mathbf{g}}_{1,k,l}^{\mathrm{H}} \mathbf{\Phi}_l \tilde{\mathbf{g}}_{2,\mu,k,l}, 
\label{eq:III.1.4}
\end{align}
which can be equivalently rewritten as
\begin{align}
y_{\mu, k} =\tilde{\mathbf{g}}_{1,k,l}^{\mathrm{H}} \mathrm{Diag}\{\tilde{\mathbf{g}}_{2,\mu,k,l}
\} \boldsymbol{\theta}_{l},
\label{eq:III.1.5}
\end{align}
where $\boldsymbol{\theta}_{l}$ is the phase shift vector of the $l$-th SIM layer extracted from the $l$-th SIM phase shift matrix $\mathbf{\Phi}_l$. By substituting the equivalent expression for the received symbol $y_{\mu, k}$, i.e.,~\eqref{eq:III.1.5} into~\eqref{eq:II.3.1}, we can rewrite the safety margin $\vert\overrightarrow{\mathrm{BC}_{\mu, k}}\vert$ as
\begin{align}
    \vert\overrightarrow{\mathrm{BC}_{\mu, k}}\vert=\Re\left\{\tilde{\mathbf{g}}_{1,k,l}^{\mathrm{H}} \mathrm{Diag}\{\tilde{\mathbf{g}}_{2,\mu,k,l}
\} \boldsymbol{\theta}_{l} e^{-j\angle s_{\mu, k}}\right\}\tan\varphi\nonumber\\
-\vert\Im\left\{\tilde{\mathbf{g}}_{1,k,l}^{\mathrm{H}} \mathrm{Diag}\{\tilde{\mathbf{g}}_{2,\mu,k,l}
\} \boldsymbol{\theta}_{l} e^{-j\angle s_{\mu, k}}\right\}\vert.\label{eq:III.1.6}
\end{align} \par

\subsubsection{Subproblem Formulation}
The expression in~\eqref{eq:III.1.6} for the safety margin of the $k$-th user during the $\mu$-th time slot depends on the phase shifts of the $l$-th SIM layer only through the phase shift vector $\boldsymbol{\theta}_{l}$. This observation allows us to decompose problem ($\mathrm{P}1$) into a series of subproblems of optimizing $\boldsymbol{\theta}_{l},\ l = 1, 2, \cdots, L$. \par

The subproblem of optimizing $\boldsymbol{\theta}_{l}$ is formulated as
\begin{subequations}
    \label{eq:III.1.7}
    \begin{align}
        (\mathrm{P}2):&\max\limits_{\boldsymbol{\theta}_{l}} \ \min\limits_{\mu,k}\ \Re\left\{\tilde{\mathbf{g}}_{1,k,l}^{\mathrm{H}} \mathrm{Diag}\{\tilde{\mathbf{g}}_{2,\mu,k,l}
\} e^{-j\angle s_{\mu,k}} \boldsymbol{\theta}_{l}\right\}\tan\varphi \nonumber\\
&\quad\quad-\vert\Im\left\{\tilde{\mathbf{g}}_{1,k,l}^{\mathrm{H}} \mathrm{Diag}\{\tilde{\mathbf{g}}_{2,\mu,k,l}
\} e^{-j\angle s_{\mu,k}} \boldsymbol{\theta}_{l}\right\}\vert \label{eq:III.1.7a}\\
        &\mathrm{s.t.}\ \vert\theta_{l,n}\vert = 1,\forall n\in\mathcal{N}.\label{eq:III.1.7b}
    \end{align}
\end{subequations}
Note that in subproblem $(\mathrm{P}2)$, $\boldsymbol{\theta}_{l}$ is the sole variable. Consequently, we can further simplify the objective of $(\mathrm{P}2)$ by defining the consolidated channel gain in its standard column-vector form $\tilde{\mathbf{g}}_{\mu,k,l}\triangleq \mathrm{Diag}\{\tilde{\mathbf{g}}_{2,\mu,k,l}
\}^{\mathrm{H}} \tilde{\mathbf{g}}_{1,k,l} \in \mathbb{C}^{N \times 1}$, leading to the reformulated $(\mathrm{P}2)$ as
\begin{subequations}
    \label{eq:III.1.8}
    \begin{align}
        (\mathrm{P}2\mathrm{a}):&\max\limits_{\boldsymbol{\theta}_{l}} \ \min\limits_{\mu,k}\ \Re\left\{\tilde{\mathbf{g}}_{\mu,k,l}^{\mathrm{H}} e^{-j\angle s_{\mu,k}} \boldsymbol{\theta}_{l}\right\}\tan\varphi \nonumber \\
        &\quad\quad\quad\quad-\vert\Im\left\{\tilde{\mathbf{g}}_{\mu,k,l}^{\mathrm{H}} e^{-j\angle s_{\mu,k}} \boldsymbol{\theta}_{l}\right\}\vert \label{eq:III.1.8a}\\
        &\mathrm{s.t.}\ ~\eqref{eq:III.1.7b}. \nonumber
    \end{align}
\end{subequations}

\subsubsection{Subproblem Identity Transformation}
Problem $(\mathrm{P}2\mathrm{a})$ is a complex-valued optimization problem subject to constant modulus constraint, with challenges arising from non-convexity and nonlinearity. To achieve tractable analysis and processing, we convert the complex-valued objective function~\eqref{eq:III.1.8a} into an equivalent real-valued formulation. \par

The absolute value operation in $(\mathrm{P}2\mathrm{a})$ is handled by decomposing objective function~\eqref{eq:III.1.8a} into two algebraic expressions, representing the positive and negative cases of the imaginary part. By defining $\tilde{\mathbf{g}}_{\mu,k,l}^{\mathrm{R}}\triangleq\Re\left\{\tilde{\mathbf{g}}_{\mu,k,l}^{\mathrm{H}} e^{-j\angle s_{\mu, k}}\right\}^{\mathrm{T}} \in \mathbb{R}^{N \times 1}$, $\tilde{\mathbf{g}}_{\mu,k,l}^{\mathrm{I}}\triangleq\Im\left\{\tilde{\mathbf{g}}_{\mu,k,l}^{\mathrm{H}} e^{-j\angle s_{\mu, k}} \right\}^{\mathrm{T}} \in \mathbb{R}^{N \times 1}$,  $\boldsymbol{\theta}_{l}^{\mathrm{R}} \triangleq \Re\left\{\boldsymbol{\theta}_{l}\right\} \in \mathbb{R}^{N \times 1}$, $\boldsymbol{\theta}_{l}^{\mathrm{I}} \triangleq \Im\left\{\boldsymbol{\theta}_{l}\right\} \in \mathbb{R}^{N \times 1}$ and $\boldsymbol{\Theta}_l\triangleq\left[\boldsymbol{\theta}_{l}^{\mathrm{R}},\boldsymbol{\theta}_{l}^{\mathrm{I}}\right]^\mathrm{T}\in\mathbb{R}^{2\times N}$, the equivalent real-valued problem for problem $(\mathrm{P}2\mathrm{a})$ is given by
\begin{subequations}
\label{eq:III.1.9}
    \begin{align}
        (&\mathrm{P}2\mathrm{b}):
        \max\limits_{\boldsymbol{\Theta}_{l}} \ \min\limits_{\mu,k}\nonumber \\
        [ &\boldsymbol{\Theta}_{l(1,:)} \left(\tilde{\mathbf{g}}_{\mu,k,l}^{\mathrm{R}}\tan\varphi - \tilde{\mathbf{g}}_{\mu,k,l}^{\mathrm{I}}\right) - \boldsymbol{\Theta}_{l(2,:)}\left(\tilde{\mathbf{g}}_{\mu,k,l}^{\mathrm{R}}+\tilde{\mathbf{g}}_{\mu,k,l}^{\mathrm{I}}\tan\varphi\right),\nonumber\\
        &\boldsymbol{\Theta}_{l(1,:)}\left(\tilde{\mathbf{g}}_{\mu,k,l}^{\mathrm{I}}+\tilde{\mathbf{g}}_{\mu,k,l}^{\mathrm{R}}\tan\varphi\right) + 
        \boldsymbol{\Theta}_{l(2,:)}\left(\tilde{\mathbf{g}}_{\mu,k,l}^{\mathrm{R}} - \tilde{\mathbf{g}}_{\mu,k,l}^{\mathrm{I}}\tan\varphi\right)], \label{eq:III.1.9a}\\
        &\mathrm{s.t.}\ \mathrm{diag}\{\boldsymbol{\Theta}_l^{\mathrm{T}}\boldsymbol{\Theta}_l\}=\mathbf{1}_{N}, \label{eq:III.1.9b}
    \end{align}
\end{subequations}
where $\mathrm{diag}\{\boldsymbol{\Theta}_l^{\mathrm{T}}\boldsymbol{\Theta}_l\}$ extracts the diagonal elements of matrix $\boldsymbol{\Theta}_l^{\mathrm{T}}\boldsymbol{\Theta}_l \in \mathbb{R}^{N \times N}$ into an all-ones column vector $\mathbf{1}_{N}$. \par

The constant modulus constraint of SIM elements in~\eqref{eq:III.1.7b} can be replaced with~\eqref{eq:III.1.9b}, where each column of $\boldsymbol{\Theta}_l$ has unit 2-norm in $\mathbb{R}^2$. Constraint \eqref{eq:III.1.9b} can be regarded as a product of $N$ circles, and forms a 2$N$-dimensional oblique manifold space $\mathcal{M}=\{\boldsymbol{\Theta}_l\in\mathbb{R}^{2\times N}:\mathrm{diag}\{\boldsymbol{\Theta}_l^{\mathrm{T}}\boldsymbol{\Theta}_l\}=\mathbf{1}_{N}\}$. Due to the smooth search space $\boldsymbol{\Theta}_l$, subproblem $(\mathrm{P}2\mathrm{a})$ can be solved in Riemannian space as an unconstrained problem~\cite{9373634}. \par

\subsection{Riemannian Optimization of Subproblem}\label{Sec.III.2}
Riemannian manifold optimization~\cite{10829726} is an effective tool to find optimal solutions to problems of the form $\min\limits_{x\in\mathcal{M}}f(x)$, where the search space $\mathcal{M}$ is a differentiable manifold exhibiting local flatness with a Riemannian structure. At each point on the manifold, a \textit{tangent space} is defined as a linear space that touches the manifold at that point, allowing linear algebra operations. In a tangent space, an \textit{inner product} can be defined to measure angles and lengths. \par

To effectively leverage the Riemannian manifold optimization for problem $(\mathrm{P}2\mathrm{b})$, we first address its max-min structure and the non-smooth objective function~\eqref{eq:III.1.9a}, and then reformulate it as a standard minimization problem on the 2$N$-dimensional oblique manifold space $\mathcal{M}$. \par

\subsubsection{Min-Max Reformulation}
For standard minimization on the oblique manifold $\mathcal{M}$, we reformulate the max-min problem $(\mathrm{P}2\mathrm{b})$ as an equivalent min-max problem, that is
\begin{subequations}
\label{eq:III.2.1}
    \begin{align}
        (\mathrm{P}3):\min\limits_{\boldsymbol{\Theta}_{l}} &\max\limits_{\mu,k}\ (g_{\mu,2k-1}, g_{\mu,2k}),\ k=1,\cdots,K\label{eq:III.2.1a}\\
        g_{\mu, 2k-1}&\triangleq\boldsymbol{\Theta}_{l(1,:)} \left(\tilde{\mathbf{g}}_{\mu,k,l}^{\mathrm{I}}-\tilde{\mathbf{g}}_{\mu,k,l}^{\mathrm{R}}\tan\varphi\right)
        \nonumber\\
        &+\boldsymbol{\Theta}_{l(2,:)}\left(\tilde{\mathbf{g}}_{\mu,k,l}^{\mathrm{R}}+\tilde{\mathbf{g}}_{\mu,k,l}^{\mathrm{I}}\tan\varphi\right),\label{eq:III.2.1b}\\
        g_{\mu, 2k}&\triangleq-\boldsymbol{\Theta}_{l(1,:)}\left(\tilde{\mathbf{g}}_{\mu,k,l}^{\mathrm{I}}+\tilde{\mathbf{g}}_{\mu,k,l}^{\mathrm{R}}\tan\varphi\right)\nonumber\\
        &+\boldsymbol{\Theta}_{l(2,:)} \left(\tilde{\mathbf{g}}_{\mu,k,l}^{\mathrm{I}}\tan\varphi-\tilde{\mathbf{g}}_{\mu,k,l}^{\mathrm{R}}\right),\label{eq:III.2.1c}\\
        \mathrm{s.t.}&\  \nonumber ~\eqref{eq:III.1.9b}.
    \end{align}
\end{subequations}
By rearranging $\{g_{\mu,2k-1}, g_{\mu,2k} \vert \quad k=1,2,\cdots,K\}$ as $\{g_{\mu,k} \vert k=1,2,\cdots,2K\}$, problem ($\mathrm{P}3$) reduces to
\begin{subequations}
    \label{eq:III.2.2}
    \begin{align}
        (\mathrm{P}3\mathrm{a}):&\min\limits_{\boldsymbol{\Theta}_{l}}\max\limits_{\mu,k}\ g_{\mu,k},\ k=1,2,\cdots,2K\label{eq:III.2.2a}\\
        &\mathrm{s.t.}\ \nonumber ~\eqref{eq:III.1.9b}.
    \end{align}
\end{subequations}
The min-max structure of the objective function~\eqref{eq:III.2.2a} in problem $(\mathrm{P}3\mathrm{a})$ is inherently non-smooth and non-differentiable. Such properties violate the smoothness requirement necessary for Riemannian manifold optimization. \par

\subsubsection{Smoothing via Log-Sum-Exp Approximation}
The non-differentiability of the $\max$ function in objective function~\eqref{eq:III.2.2a} is challenging for Riemannian optimization. To overcome this, we employ the smooth log-sum-exp approximation to transform the non-smooth $\max$ function $\max\limits_{\mu,k}\ g_{\mu,k},\ k=1,2,\cdots,2K$, into a differentiable form, leading to the following smoothed problem $(\mathrm{P}4)$
\begin{subequations}
    \label{eq:III.2.3}
    \begin{align}
        (\mathrm{P}4):&\min\limits_{\boldsymbol{\Theta}_{l}}\ 
        f(\boldsymbol{\Theta}_{l})=\varepsilon\log\left(\sum_{\mu=1}^{U}\sum_{k=1}^{2K}\exp\left(g_{\mu,k}/\varepsilon\right)\right)
        \label{eq:III.2.3a}\\
        &\mathrm{s.t.}\ \nonumber ~\eqref{eq:III.1.9b},
    \end{align}
\end{subequations}
where $\varepsilon > 0$ is a constant to balance numerical stability and approximation accuracy. 

\begin{algorithm}[t]
\caption{ROM-Based SIM Phase Shift Optimization}
\renewcommand{\algorithmicrequire}{\textbf{Input:}}
\renewcommand{\algorithmicensure}{\textbf{Output:}} 
\newcommand{\algorithmicinitialize}{\textbf{Initialization:}}
\newcommand{\INITIALIZE}{\item[\algorithmicinitialize]}
\begin{algorithmic}[1]
    \REQUIRE CSI $\mathbf{H}$ and $\{\mathbf{Q}_l\}_{l=1}^{L}$; symbol vector $\{\mathbf{s}_{\mu}\}_{\mu = 1}^{U}$; number of iterations $T$; number of layers $L$.
    \INITIALIZE
    Randomly initialize $\{\mathbf{\Phi}_{l}\}_{l=1}^L$.
    \FOR{Iteration $t=1,2,\cdots,T$}
        \FOR{Step $l=L,L-1,\cdots,1$}
        \STATE Obtain $\mathbf{\Phi}_{l}$ and $\boldsymbol{\theta}_{l}=\mathrm{diag} \left\{ \mathbf{\Phi}_{l} \right\}$;
        \FOR{Slot $\mu = 1, 2, \cdots, U$}
            \STATE Update $\tilde{\mathbf{g}}_{1,k,l}$ and $\tilde{\mathbf{g}}_{2,\mu,k,l}$ by~\eqref{eq:III.1.2} and~\eqref{eq:III.1.3};
            \STATE Update $\tilde{\mathbf{g}}_{\mu,k,l}$ by $\tilde{\mathbf{g}}_{\mu,k,l} = \mathrm{Diag}\{\tilde{\mathbf{g}}_{2,\mu,k,l}
    \}^{\mathrm{H}} \tilde{\mathbf{g}}_{1,k,l}$;
            \STATE Divide the complex-valued $\boldsymbol{\theta}_{l}$ and $\tilde{\mathbf{g}}_{\mu,k,l}$ into real-valued $\tilde{\mathbf{g}}_{\mu,k,l}^{\mathrm{R}}$, $\tilde{\mathbf{g}}_{\mu,k,l}^{\mathrm{I}}$, $\boldsymbol{\theta}_{l}^{\mathrm{R}}$, $\boldsymbol{\theta}_{l}^{\mathrm{I}}$;
            \STATE Construct $\boldsymbol{\Theta}_l$ by $\boldsymbol{\Theta}_l = \left[\boldsymbol{\theta}_{l}^{\mathrm{R}},\boldsymbol{\theta}_{l}^{\mathrm{I}}\right]^\mathrm{T}$;
            \STATE Update $g_{\mu, 2k-1},g_{\mu, 2k},k=1,2,\cdots,K$ by~\eqref{eq:III.2.1b} and~\eqref{eq:III.2.1c};
            \STATE Substitute $g_{\mu, k},k=1,2,\cdots,2K$ into~\eqref{eq:III.2.3};
        \ENDFOR
        \STATE Update $\boldsymbol{\theta}_{l}$ by solving $(\mathrm{P}4)$;
        \STATE Update $\mathbf{\Phi}_{l}$ by $\mathbf{\Phi}_l=\mathrm{Diag} \left\{ \boldsymbol{\theta}_{l} \right\}$.
    \ENDFOR
    \IF{$(\vert\overrightarrow{\mathrm{BC}}\vert_{\min}^{t}-\vert\overrightarrow{\mathrm{BC}}\vert_{\min}^{t-1})
/\vert\overrightarrow{\mathrm{BC}}\vert_{\min}^{t-1} \leq \epsilon$}
    \STATE Break;
    \ENDIF
    \ENDFOR
    \ENSURE $\{\mathbf{\Phi}_{l,\mathrm{opt}}\}_{l=1}^L$.
\end{algorithmic}
\label{alg:1}
\end{algorithm}

\begin{lemm}
    The smooth log-sum-exp approximation $f(\boldsymbol{\Theta}_{l})$ is a smooth maximum approximation function by leveraging the properties of the exponential and logarithmic functions. It satisfies $f(\boldsymbol{\Theta}_{l}) \geq g_{\mu^*,k^*}$, and $f(\boldsymbol{\Theta}_{l}) \to g_{\mu^*,k^*}$ when $\varepsilon \to 0$, where $g_{\mu^*,k^*}=\max\limits_{\mu,k}\ g_{\mu,k}$.
\end{lemm}

\begin{IEEEproof}
We begin by extracting the maximum term $g_{\mu*,k*}$ from the summation:
\begin{subequations}
\label{eq:lemm.1}
\begin{align}
    f(\boldsymbol{\Theta}_l) &\!=\! \varepsilon \log\! \Big(\!\! \exp\left(g_{\mu^*,k^*}/\varepsilon\right) 
    \sum_{\mu\!, k} \!\exp\!\left(g_{\mu,k} \!- \!g_{\mu^*\!,k^*}/\varepsilon \right)\!\! \Big)\! \label{eq:lemm.1a}\\
    &=g_{\mu^*,k^*} + \varepsilon \log \Big( \sum_{\mu, k}\exp(g_{\mu,k} - g_{\mu^*,k^*}/\varepsilon) \Big) \label{eq:lemm.1b} \\
    &\geq g_{\mu^*,k^*}. \label{eq:lemm.1c}
\end{align}
\end{subequations}
Since $g_{\mu,k} \leq g_{\mu^*,k^*}$ for all $\mu, k$ and $\varepsilon \to 0$, the second term of~\eqref{eq:lemm.1b} approaches 0. Consequently, $f(\boldsymbol{\Theta}_{l}) \to g_{\mu^*,k^*}$. It holds that $f(\boldsymbol{\Theta}_{l})=g_{\mu^*,k^*}$ if and only if the input sequence $\{g_{\mu,k}\}_{\mu = 1, \cdots, U, \ k = 1, \cdots, K}$ consists of a single unique element. Therefore, $f(\boldsymbol{\Theta}_{l})$ serves as an upper bound for $\max\limits_{\mu,k}\ g_{\mu,k}$ ensuring the equivalence between problem~$(\mathrm{P}3\mathrm{a})$ and problem~$(\mathrm{P}4)$.
\end{IEEEproof}

\subsubsection{Riemannian Optimization Procedure and Iterative Update}
With problem $(\mathrm{P}4)$ now cast as the minimization of a smooth objective function~\eqref{eq:III.2.3a} over a 2$N$-dimensional oblique manifold $\mathcal{M}$ defined by~\eqref{eq:III.1.9b}, it becomes a standard problem that can be efficiently solved via Riemannian optimization~\cite{boumal2014manopt}.
The subsequent procedure involves executing the iterative optimization algorithm on this manifold, which requires the computation of key geometric objects, such as Euclidean gradient and its projection onto the tangent space, followed by the refraction mapping operation to converge to a solution.\par

To apply the Riemannian optimization to $(\mathrm{P}4)$, we first define a search space $\mathcal{M}=\{\boldsymbol{\Theta}_l\in\mathbb{R}^{2\times N}:\mathrm{diag}\{\boldsymbol{\Theta}_l^{\mathrm{T}}\boldsymbol{\Theta}_l\}=\mathbf{1}_{N}\}$ , which constitutes a $2N$-dimensional oblique manifold, namely the Cartesian product of $N$ unit circles. The tangent space $\mathcal{T}_{\boldsymbol{\Theta}_l^{i}}\mathcal{M}$ at the $i$-th iteration point $\boldsymbol{\Theta}_l^{i}$ is given by
\begin{align}
    \mathcal{T}_{\boldsymbol{\Theta}_l^{i}}\mathcal{M}=\left\{\mathbf{Z}_{\boldsymbol{\Theta}_l^{i}}\in\mathbb{R}^{2\times
     N}:\mathrm{diag}\left(\boldsymbol{\Theta}_l^{\mathrm{T}}\mathbf{Z}_{\boldsymbol{\Theta}_l^{i}}\right)=\mathbf{0}\right\}.\label{eq:III.2.4}
\end{align} \par

\begin{figure*}[!b]
\normalsize
\begin{subequations}
\hrulefill
\vspace*{4pt}
\label{eq:III.2.5}
    \begin{align}
        \nabla f\left(\boldsymbol{\Theta}_{l(1,:)}^{i}\right)=\frac{\sum_{\mu=1}^{U}\sum_{k=1}^{K}\left\{\exp(g_{\mu, 2k-1}/\varepsilon)\left(\tilde{\mathbf{g}}_{\mu,k,l}^{\mathrm{I}}-\tilde{\mathbf{g}}_{\mu,k,l}^{\mathrm{R}}\tan\varphi\right)^{\mathrm{T}}+\exp(g_{\mu, 2k}/\varepsilon)\left[-\left(\tilde{\mathbf{g}}_{\mu,k,l}^{\mathrm{I}}+\tilde{\mathbf{g}}_{\mu,k,l}^{\mathrm{R}}\tan\varphi\right)^{\mathrm{T}}\right]\right\}}{\sum_{\mu=1}^{U}\sum_{k=1}^{2K}\exp\left(g_{\mu, k}/\varepsilon\right)},\label{eq:III.2.5a}\\
        \nabla f\left(\boldsymbol{\Theta}_{l(2,:)}^{i}\right)=\frac{\sum_{\mu=1}^{U}\sum_{k=1}^{K}\left\{\exp(g_{\mu, 2k-1}/\varepsilon)\left(\tilde{\mathbf{g}}_{\mu,k,l}^{\mathrm{R}}+\tilde{\mathbf{g}}_{\mu,k,l}^{\mathrm{I}}\tan\varphi\right)^{\mathrm{T}}+\exp(g_{\mu, 2k}/\varepsilon)\left(\tilde{\mathbf{g}}_{\mu,k,l}^{\mathrm{I}}\tan\varphi-\tilde{\mathbf{g}}_{\mu,k,l}^{\mathrm{R}}\right)^{\mathrm{T}}\right\}}{\sum_{\mu=1}^{U}\sum_{k=1}^{2K}\exp\left(g_{\mu, k}/\varepsilon\right)}.\label{eq:III.2.5b}
    \end{align}
\end{subequations}
\end{figure*}

With the manifold and its tangent spaced defined, the next step is to compute the direction of gradient descent on the manifold. This begins with the calculation of the Euclidean gradient of the objection function~\eqref{eq:III.2.3a}. The Euclidean gradient of $f\left(\boldsymbol{\Theta}_{l}^{i}\right)$ at the $i$-th iteration point $\boldsymbol{\Theta}_l^{i}$, i.e., $\nabla f\left(\boldsymbol{\Theta}_{l}^{i}\right)$, is given in~\eqref{eq:III.2.5} (at the bottom of the next page). \par

As the Euclidean gradient $\nabla f\left(\boldsymbol{\Theta}_{l}^{i}\right)$ does not lie in the tangent space of the manifold $\mathcal{T}_{\boldsymbol{\Theta}_l^{i}}\mathcal{M}$, the associated Riemannian gradient $\nabla_{\mathcal{M}} f\left(\boldsymbol{\Theta}_{l}^{i}\right)$ is obtained by orthogonally projecting $\nabla f\left(\boldsymbol{\Theta}_{l}^{i}\right)$ onto $\mathcal{T}_{\boldsymbol{\Theta}_l^{i}}\mathcal{M}$, as given by
\begin{align}
    \nabla_{\mathcal{M}} f\left(\boldsymbol{\Theta}_{l}^{i}\right)&=\mathrm{Proj}_{\mathcal{T}_{\boldsymbol{\Theta}_l^{i}}\mathcal{M}}\left(\nabla f\left(\boldsymbol{\Theta}_{l}^{i}\right)\right)\nonumber\\
    &\!=\!\nabla \!f\left(\boldsymbol{\Theta}_{l}^{i}\right)\!-\!\boldsymbol{\Theta}_{l}^{i}\mathrm{Diag}\left\{\left[\boldsymbol{\Theta}_{l}^{i}\right]^{\mathrm{T}}\nabla \!f\left(\boldsymbol{\Theta}_{l}^{i}\right)\right\},\label{eq:III.2.6}
\end{align}
where $\mathrm{Proj}\{\cdot\}$ denotes the projection operator. \par

Equipped with the Riemannian gradient $\nabla_{\mathcal{M}} f\left(\boldsymbol{\Theta}_{l}^{i}\right)$, we can now update the current point $\boldsymbol{\Theta}_{l}^{i}$ by moving along the negative gradient direction within the tangent space, analogous to the procedure in Euclidean gradient descent:
\begin{align}
    \tilde{\boldsymbol{\Theta}}_{l}^{i}=\boldsymbol{\Theta}_{l}^{i}-\zeta\nabla_{\mathcal{M}} f\left(\boldsymbol{\Theta}_{l}^{i}\right),
    \label{eq:III.2.7}
\end{align}
where $\zeta$ is the step size to ensure stability and convergence of the algorithm. \par 

However, the updated point $\tilde{\boldsymbol{\Theta}}_{l}^{i}$ is still located in the tangent space $\mathcal{T}_{\boldsymbol{\Theta}_l^{i}}\mathcal{M}$, instead of the oblique manifold space $\mathcal{M}$. 
A~\textit{retraction} mapping operation is adopted to map the updated point $\tilde{\boldsymbol{\Theta}}_{l}^{i}$ from the tangent space $\mathcal{T}_{\boldsymbol{\Theta}_l^{i}}\mathcal{M}$ onto the manifold $\mathcal{M}$ to obtain the next point $\boldsymbol{\Theta}_{l}^{i+1}$, as given by
\begin{align}
    \boldsymbol{\Theta}_{l}^{i+1}=\frac{\tilde{\boldsymbol{\Theta}}_{l}^{i}}{\Vert\tilde{\boldsymbol{\Theta}}_{l}^{i}\Vert},
    \label{eq:III.2.8}
\end{align}
where $\Vert\cdot\Vert$ is the generalized Euclidean norm. The phase shift $\boldsymbol{\theta}_{l}^{*}$ is obtained as $\boldsymbol{\theta}_{l}^{*}=\left[\boldsymbol{\Theta}_{l(1,:)}^{*}\right]^{\mathrm{T}}+j\left[\boldsymbol{\Theta}_{l(2,:)}^{*}\right]^{\mathrm{T}}$ with the minimum objective value $f(\boldsymbol{\Theta}_{l}^{*})$.

\subsection{Proposed Recursive Oblique Manifold Algorithm}\label{Sec.III.3}
This subsection presents the complete ROM-based SIM phase shift optimization, as summarized in Alg.~\ref{alg:1}. The proposed ROM algorithm operates through an iterative, layer-by-layer optimization strategy to address the variable coupling across different SIM layers. \par
 
The ROM algorithm is developed to optimize phase shifts $\mathbf{\Phi}_l$ layer-by-layer.
It proceeds over a maximum of $T$ iterations. Within each iteration $t$, there are $L$ sequential steps optimizing the phase shifts of each layer from $l = L$ to $l = 1$.
In the $l$-th step, only the $l$-th layer phase shifts $\mathbf{\Phi}_l$ are optimized, which corresponds to solving subproblem $(\mathrm{P}4)$ on the oblique manifold using the Riemannian procedure detailed in Section~\ref{Sec.III.2}, while the phase shifts of the other layers remain fixed. \par

Following the optimization of phase shift matrices across $L$ layers in the $t$-th iteration, the updated $\{\mathbf{\Phi}_{l,t}\}_{l=1}^L$ are incorporated into~\eqref{eq:II.1.2.2} and~\eqref{eq:II.3.1} to evaluate the minimum safety margin $|\overrightarrow{\mathrm{BC}}|_{\min}^{t}$ across all users. The ROM algorithm terminates when either the maximum number of iterations $T$ is reached, or the absolute difference between the minimum safety margin among all users obtained between adjacent iterations is less than the tolerant error $\epsilon$, i.e., $(\vert\overrightarrow{\mathrm{BC}}\vert_{\min}^{t}-\vert\overrightarrow{\mathrm{BC}}\vert_{\min}^{t-1})
/\vert\overrightarrow{\mathrm{BC}}\vert_{\min}^{t-1} \leq \epsilon$. \par

\begin{algorithm}[t]
\caption{Greedy Safety Margin-Based AS Algorithm}
\renewcommand{\algorithmicrequire}{\textbf{Input:}}
\renewcommand{\algorithmicensure}{\textbf{Output:}} 
\newcommand{\algorithmicinitialize}{\textbf{Initialization:}}
\newcommand{\INITIALIZE}{\item[\algorithmicinitialize]}
\begin{algorithmic}[1]
    \REQUIRE $\mathbf{H}$, $\{\mathbf{Q}_l\}_{l=2}^{L}$, $\{\mathbf{s}_{\mu}\}_{\mu = 1}^{U}$, $\{\mathbf{q}_i\}_{i=1}^{N_m}$, $\mathcal{P}_0 = \varnothing$, $\{\mathbf{\Phi}_{l}\}_{l=1}^L$.
    \FOR{User $k = 1,2,\cdots,K$}
        \FOR{Antenna $m = 1,2,\cdots,M$}
            \IF{$m \notin \mathcal{P}_{k-1}$}
                \STATE Add the $m$-th antenna index into the AS set, i.e., $\mathcal{P}_{k}^{\prime} = \mathcal{P}_{k-1} \cup {m}$;
                \FOR{slot $\mu = 1, 2, \cdots, U$}
                    \STATE Calculate $\vert\overrightarrow{\mathrm{BC'}_{\mu, k, \mathcal{P}_k^{\prime}}}\vert$ by using~\eqref{eq:IV.0.2};
                \ENDFOR
            \ENDIF
        \ENDFOR
        \STATE Find the maximum $\min\limits_{\mu, k}\vert\overrightarrow{\mathrm{BC'}_{\mu, k, \mathcal{P}_{k}^{\prime}}}\vert$ and its associated antenna $m^{*}$ as the $m$-th antenna;
        \STATE $\mathcal{P}_{k} = \mathcal{P}_{k-1} \cup {m^{*}}$;
        \STATE Perform $\tilde{\mathbf{Q}}_{1,\mathcal{P}_{K}(:, k)} = \mathbf{q}_{m^{*}}$;
        
    \ENDFOR
    \ENSURE AS set $\mathcal{P}_K$, and channel $\tilde{\mathbf{Q}}_{1,\mathcal{P}_K}$.
\end{algorithmic}
\label{alg:2}
\end{algorithm}

\section{Antenna Selection and Power Allocation}\label{Sec.IV}
In practice, the number of served users $K$ varies dynamically, while the BS is equipped with a tailored configuration of $M>K$ antennas collected by $\mathcal{N}_M=\{1,\cdots,M\}$. To acquire the diversity benefits with reduced hardware costs, the AS becomes indispensable for MIMO systems. Such selection not only reduces the use of RF chains, but also enables better utilization of MUI to further reduce the SER. On the other hand, the AS also requires PA to achieve the proper eigen-channel assignment under the total power budget.

This section considers the AS and PA in the SIM-enabled IEP system.
Assume that $K$ antennas are selected from total $M$ antennas denoted by $\mathcal{P}_K = \{1,2,\cdots, K\}$, to serve $K$ users. In the considered AS scenario, the received signal at the $k$-th user in the $\mu$-th time slot within a frame served by the selected antennas $\mathcal{P}_K$ is given by
\begin{align}
\tilde{y}_{\mu, k, \mathcal{P}_K} = \mathbf{h}_k^{\mathrm{H}} \mathbf{G} \tilde{\mathbf{Q}}_{1,\mathcal{P}_K} \mathbf{P} \mathbf{s}_{\mu},
\label{eq:IV.0.1}
\end{align}
where $\tilde{\mathbf{Q}}_{1,\mathcal{P}_K} \in \mathbb{C}^{N \times K}$ denotes the channel between the first SIM layer and the selected transmit antennas $\mathcal{P}_K$.
$\mathbf{P} = \mathrm{Diag} \left\{ \sqrt{p_1}, \sqrt{p_2}, \cdots, \sqrt{p_K} \right\}$ denotes the PA matrix, and F-norm of $\mathbf{P}$ is constrained to 1. At present, the safety margin $\vert\overrightarrow{\mathrm{BC'}_{\mu, k, \mathcal{P}_K}}\vert$ of the $k$-th user in the $\mu$-th time slot served by the selected antennas $\mathcal{P}_K$ is denotes as
\begin{align}
    \vert\overrightarrow{\mathrm{BC'}_{\mu, k, \mathcal{P}_K}}\vert&=\Re\left\{\tilde{y}_{\mu, k, \mathcal{P}_K}e^{-j\angle s_{\mu, k}}\right\}\tan\varphi \nonumber \\ 
    &-\vert\Im\left\{\tilde{y}_{\mu, k, \mathcal{P}_K}e^{-j\angle s_{\mu, k}}\right\}\vert.\label{eq:IV.0.2}
\end{align}

We jointly optimize AS, SIM phase shifts $\{\mathbf{\Phi}_{l,\mathrm{opt}}\}_{l=1}^L$, and PA matrix to maximize the minimum safety margin of users within a coherent frame. The problem can be formulated as
\begin{subequations}
\label{eq:IV.0.3}
    \begin{align}
        (\mathrm{P}5):&\max\limits_{\mathcal{P}_K, \{\mathbf{\Phi}_l\}_{l=1}^L, \mathbf{P}}\ \min\limits_{\mu,k}\ \vert\overrightarrow{\mathrm{BC'}_{\mu, k, \mathcal{P}_K}}\vert \label{eq:IV.0.3a}\\
        &\mathrm{s.t.}\ \mathcal{P}_K \subseteq \mathcal{N}_M \label{eq:IV.0.3b} \\
        &\quad \quad \vert\theta_{l,n}\vert = 1,\forall n\in\mathcal{N},\forall l\in\mathcal{L} \label{eq:IV.0.3c} \\
        &\quad \quad \Vert \mathbf{P} \Vert_F = 1. \label{eq:IV.0.3d}       
    \end{align}
\end{subequations}
Similarly, substituting $\vert\overrightarrow{\mathrm{BC'}_{\mu, k, \mathcal{P}_K}}\vert$ in \eqref{eq:IV.0.2} into problem $(\mathrm{P}5)$ yields
\begin{subequations}
\label{eq:IV.0.4}
    \begin{align}
        (\mathrm{P}5\mathrm{a}):&\max\limits_{\mathcal{P}_K, \{\mathbf{\Phi}_l\}_{l=1}^L, \mathbf{P}} \min\limits_{\mu,k}\Re\left\{\tilde{y}_{\mu, k, \mathcal{P}_K} e^{-j\angle s_{\mu,k}}\right\}\tan\varphi \nonumber\\
        &\quad\quad\quad\quad\quad\quad\quad-\vert\Im\left\{\tilde{y}_{\mu, k, \mathcal{P}_K} e^{-j\angle s_{\mu,k}}\right\}\vert\label{eq:IV.0.4a} \\
        &\mathrm{s.t.}\  ~\eqref{eq:IV.0.3b}, ~\eqref{eq:IV.0.3c}, ~\eqref{eq:IV.0.3d}. \nonumber
    \end{align}
\end{subequations}
Analogous to problem ($\mathrm{P}1$), problem ($\mathrm{P}5\mathrm{a}$) is also non-convex due to the objective function in~\eqref{eq:IV.0.4a}, coupled non-convex variables within the constant modulus constraint in~\eqref{eq:IV.0.3c} and unit energy constraint in~\eqref{eq:IV.0.3d}, and the discontinuous solution space~\eqref{eq:IV.0.3b} for transmit AS.

In this paper, we develop an ROM-AO framework to tackle the non-convex problem ($\mathrm{P}5\mathrm{a}$). In particular, the problem ($\mathrm{P}5\mathrm{a}$) is decoupled into three subproblems by using the AO framework, including transmit AS, SIM phase shift optimization, and PA. Given SIM phase shifts and PA matrix, we firstly select the antennas $\mathcal{P}_K$ by the proposed greedy safety margin-based AS algorithm; given the selected transmit antennas and PA matrix, we optimize the SIM phase shifts $\{\mathbf{\Phi}_{l,\mathrm{opt}}\}_{l=1}^L$ layer-by-layer through the ROM algorithm proposed in Section~\ref{Sec.III}; given the selected transmit antennas and SIM phase shifts, we find the PA policy to further push the constellation points towards the CI region under the fixed total power budget.

\subsection{Greedy Safety Margin-Based Transmit Antenna Selection}\label{Sec.IV.1}
Given the SIM phase shifts and PA matrix, $K$ transmit antennas are selected from $M$ antennas to guarantee the quality of service (QoS) of the served users. The AS problem is formulated as 
\begin{subequations}
    \label{eq:IV.1.1}
    \begin{align}
    (\mathrm{P}6):&\max\limits_{\mathcal{P}_K}\ \min\limits_{\mu,k}\ \vert\overrightarrow{\mathrm{BC'}_{\mu, k, \mathcal{P}_K}}\vert \label{eq:IV.1.1.1} \\
    &\mathrm{s.t.}\ \mathcal{P}_K \subseteq \mathcal{N}_M, \label{eq:IV.1.1.2}
    \end{align}
\end{subequations}
which is a combinatorial problem. 
To avoid the substantial complexity of exhaustive search, we propose a greedy safety margin-based AS algorithm to solve $(\mathrm{P}6)$. Specifically, the AS begins from the empty set $\mathcal{P}_0 = \varnothing$, expand to the final set $\mathcal{P}_K$, as antenna $m \notin \mathcal{P}_{k-1}$ is incrementally added. When selecting the $m$-th antenna and combining it with the previous selected antennas $\mathcal{P}_{k-1}$ to jointly serve $k$ users, the antenna that can achieve the maximum $\min\limits_{\mu, k}\vert\overrightarrow{\mathrm{BC'}_{\mu, k, \mathcal{P}_{k}^{\prime}}}\vert$ is selected.
Correspondingly, the number of served users grows from the first user to a total of $K$ users, until the cardinality of the set reaches $K$.

The greedy safety margin-based AS algorithm is summarized in Algorithm~\ref{alg:2}, where $\mathbf{q}_m \in \mathbb{C}^{N \times 1}$ represents the channel between the $m$-th antenna to the first SIM layer, and the channels between the first SIM layer and the selected antennas construct the channel matrix $\tilde{\mathbf{Q}}_{1,\mathcal{P}_K}$.

\subsection{Oblique Manifold Algorithm for Power Allocation}\label{Sec.IV.2}
Given the SIM phase shifts and the selected transmit antennas, the PA subproblem can be formulated as
\begin{subequations}
    \label{eq:IV.2.1}
    \begin{align}
    (\mathrm{P}7):&\max\limits_{\mathbf{p}}\ \min\limits_{\mu,k}\ \vert\overrightarrow{\mathrm{BC'}_{\mu, k, \mathcal{P}_K}}\vert \label{eq:IV.2.1a} \\
    &\mathrm{s.t.}\ \Vert \mathbf{p} \Vert_2 = 1, \label{eq:IV.2.1b}
    \end{align}
\end{subequations}
where $\mathbf{p} \triangleq \mathrm{diag}\{ \mathbf{P} \} \in \mathbb{R}^{K \times 1}$ denotes the PA vector. Problem $(\mathrm{P}7)$ is non-convex due to the discontinuous and non-differentiability objective function in~\eqref{eq:IV.2.1a} and the unit energy constraint in~\eqref{eq:IV.2.1b}, and the Riemannian manifold optimization is adopted to tackle this non-convex problem.

Initially, we rewrite the received signal~\eqref{eq:II.1.2.2} as
\begin{align}
    \tilde{y}_{\mu, k, \mathcal{P}_K} = \mathbf{h}_k^{\mathrm{H}} \mathbf{G} \tilde{\mathbf{Q}}_{1,\mathcal{P}_K} \mathrm{Diag}\{\mathbf{s}_{\mu}\}\mathbf{p}. \label{eq:IV.2.2}
\end{align}
To simplify the expression of~\eqref{eq:IV.0.2}, we define $\tilde{\mathbf{h}}_{\mu, k, \mathcal{P}_K} \triangleq \left(\mathbf{h}_k^{\mathrm{H}} \mathbf{G} \tilde{\mathbf{Q}}_{1,\mathcal{P}_K} \mathrm{Diag}\{\mathbf{s}_{\mu}\} e^{-j\angle s_{\mu, k}}\right)^{\mathrm{H}} \in \mathbb{C}^{K \times 1}$, $\tilde{\mathbf{h}}_{\mu, k, \mathcal{P}_K}^{\mathrm{R}} \triangleq \Re\left\{\tilde{\mathbf{h}}_{\mu, k, \mathcal{P}_K}^{\mathrm{H}}\right\}^{\mathrm{T}} \in \mathbb{R}^{K \times 1}$ and $\tilde{\mathbf{h}}_{\mu, k, \mathcal{P}_K}^{\mathrm{I}} \triangleq \Im\left\{\tilde{\mathbf{h}}_{\mu, k, \mathcal{P}_K}^{\mathrm{H}}\right\}^{\mathrm{T}} \in \mathbb{R}^{K \times 1}$. $\vert\overrightarrow{\mathrm{BC'}_{\mu, k, \mathcal{P}_K}}\vert$ can be rewritten as
\begin{align}
    \vert\overrightarrow{\mathrm{BC'}_{\mu, k, \mathcal{P}_K}}\vert 
 = \tilde{\mathbf{h}}_{\mu, k, \mathcal{P}_K}^{\mathrm{R},\mathrm{T}} \mathbf{p} \tan\varphi - \vert \tilde{\mathbf{h}}_{\mu, k, \mathcal{P}_K}^{\mathrm{I},\mathrm{T}} \mathbf{p} \vert.\label{eq:IV.2.3}
\end{align}

By using~\eqref{eq:IV.2.3}, the PA subproblem can be formulated as
\begin{subequations}
    \label{eq:IV.2.4}
    \begin{align}
    (\mathrm{P}7\mathrm{a}):&\max\limits_{\mathbf{p}}\ \min\limits_{\mu,k}\ \tilde{\mathbf{h}}_{\mu, k, \mathcal{P}_K}^{\mathrm{R},\mathrm{T}} \mathbf{p} \tan\varphi - \vert \tilde{\mathbf{h}}_{\mu, k, \mathcal{P}_K}^{\mathrm{I},\mathrm{T}} \mathbf{p} \vert \label{eq:IV.2.4a} \\
    &\mathrm{s.t.}\ ~\eqref{eq:IV.2.1b} \nonumber.
    \end{align}
\end{subequations}
Note that, the unit 2-norm in $\mathbb{R} ^ K$ of the constraint~\eqref{eq:IV.2.1b} constitutes a $K$-dimensional oblique manifold space. Thus, the subproblem $(\mathrm{P}7\mathrm{a})$ can be tackled in Riemannian space as an unconstrained problem and can be converted as
\begin{subequations}
    \label{eq:IV.2.5}
    \begin{align}
    (\mathrm{P}8):&\min\limits_{\mathbf{p}}\ \max\limits_{\mu,k}\ \vert \tilde{\mathbf{h}}_{\mu, k, \mathcal{P}_K}^{\mathrm{I},\mathrm{T}} \mathbf{p} \vert-\tilde{\mathbf{h}}_{\mu, k, \mathcal{P}_K}^{\mathrm{R},\mathrm{T}} \mathbf{p} \tan\varphi \label{eq:IV.2.5a} \\
    &\mathrm{s.t.}\ ~\eqref{eq:IV.2.1b} \nonumber.
    \end{align}
\end{subequations}
To convert the complex-valued objective~\eqref{eq:IV.2.5a} in problem $(\mathrm{P}8)$ into the real-valued form, we first perform some manipulations, as given by
\begin{align}
\tilde{p}_{\mu, 2k - 1, \mathcal{P}_K} &\triangleq \left( \tilde{\mathbf{h}}_{\mu, k, \mathcal{P}_K}^{\mathrm{I},\mathrm{T}} - \tilde{\mathbf{h}}_{\mu, k, \mathcal{P}_K}^{\mathrm{R},\mathrm{T}} \tan\varphi \right) \mathbf{p}, \label{eq:IV.2.6} \\
\tilde{p}_{\mu, 2k, \mathcal{P}_K} &\triangleq - \left( \tilde{\mathbf{h}}_{\mu, k, \mathcal{P}_K}^{\mathrm{I},\mathrm{T}} + \tilde{\mathbf{h}}_{\mu, k, \mathcal{P}_K}^{\mathrm{R},\mathrm{T}} \tan\varphi \right) \mathbf{p}. \label{eq:IV.2.7}
\end{align}
Afterwards, subproblem $(\mathrm{P}8)$ can be relaxed by the log-sum-exp approximation as
\begin{subequations}
    \label{eq:IV.2.8}
    \begin{align}
        (\mathrm{P}9):&\min\limits_{\mathbf{p}} 
        p(\mathbf{p})\!=\!\varepsilon_p \log\left(\sum_{\mu=1}^{U}\sum_{k=1}^{2K}\exp\left(\tilde{p}_{\mu, k, \mathcal{P}_K}/\varepsilon_p\right)\right)
        \label{eq:IV.2.8a}\\
        &\mathrm{s.t.}\ ~\eqref{eq:IV.2.1b} \nonumber.
    \end{align}
\end{subequations}
\begin{figure*}[!b]
\normalsize
\hrulefill
\vspace*{2pt}
\begin{align}
    \label{eq:IV.2.9}
    \nabla p(\mathbf{p}) = \frac{\sum_{\mu=1}^{U}\sum_{k=1}^{K}\left\{\exp(\tilde{p}_{\mu, 2k - 1, \mathcal{P}_K}/\varepsilon_p)\left(\tilde{\mathbf{h}}_{\mu, k, \mathcal{P}_K}^{\mathrm{I}}-\tilde{\mathbf{h}}_{\mu, k, \mathcal{P}_K}^{\mathrm{R}}\tan\varphi\right)+\exp(\tilde{p}_{\mu, 2k, \mathcal{P}_K}/\varepsilon_p)\left[-\left(\tilde{\mathbf{h}}_{\mu, k, \mathcal{P}_K}^{\mathrm{I}}+\tilde{\mathbf{h}}_{\mu, k, \mathcal{P}_K}^{\mathrm{R}}\tan\varphi\right)\right]\right\}}{\sum_{\mu=1}^{U}\sum_{k=1}^{2K}\exp\left(\tilde{p}_{\mu, k, \mathcal{P}_K}/\varepsilon_p\right)}.
\end{align}
\end{figure*}
The Euclidean gradient of $p(\mathbf{p})$ is given in~\eqref{eq:IV.2.9}.
Due to the unit F-norm in $\mathbb{R} ^ K$ of~\eqref{eq:IV.2.1b}, subproblem $(\mathrm{P}9)$ can be solved using the oblique manifold algorithm. The PA matrix can be obtained from the suboptimal $\mathbf{p}^{*}$, i.e., $\mathbf{P}^{*} = \mathrm{Diag}\{ \mathbf{p}^{*} \}$.

\begin{algorithm}[t]
\caption{ROM-AO Algorithm for EM Domain-Based Frame-Level IEP}
\renewcommand{\algorithmicrequire}{\textbf{Input:}}
\renewcommand{\algorithmicensure}{\textbf{Output:}} 
\newcommand{\algorithmicinitialize}{\textbf{Initialization:}}
\newcommand{\INITIALIZE}{\item[\algorithmicinitialize]}
\begin{algorithmic}[1]
    \REQUIRE $\mathbf{H}$, $\{\mathbf{Q}_l\}_{l=2}^{L}$, $\{\mathbf{s}_{\mu}\}_{\mu = 1}^{U}$, $[\mathbf{q}_1, \cdots, \mathbf{q}_{N_m}]$, $\mathcal{P}_0 = \varnothing$, $T$, $L$.
    \INITIALIZE
    Randomly initialize $\{\mathbf{\Phi}_{l}\}_{l=1}^L$ and $\mathbf{P}$.
    \FOR{$t = 1,2,\cdots,T$}
        \STATE Given $\{\mathbf{\Phi}_{l}\}_{l=1}^L$ and $\mathbf{P}$, update $\mathcal{P}_K$ and $\tilde{\mathbf{Q}}_{1,\mathcal{P}_K}$;
        \FOR{$l = 1,2,\cdots,L$}
            \STATE Given $\mathcal{P}_K$ and $\mathbf{P}$, update $\mathbf{\Phi}_{l}$ by Algorithm~\ref{alg:1};
            \STATE Update the phase shift $\mathbf{\Phi}_{l}^*$;
        \ENDFOR
        \STATE Given $\mathcal{P}_K$ and $\{\mathbf{\Phi}_{l}^*\}_{l=1}^L$, update $\mathbf{p}^{*}_{\mathrm{av}}$ by solving~\eqref{eq:IV.2.8};
        
    \ENDFOR
    \ENSURE $\mathcal{P}_K$, $\{\mathbf{\Phi}_{l,\mathrm{opt}}\}_{l=1}^L$, and $\mathbf{P}_{\mathrm{opt}}$.
\end{algorithmic}
\label{alg:3}
\end{algorithm}

The proposed ROM-AO framework decouples the non-convex problem ($\mathrm{P}5\mathrm{a}$) into three subproblems, including transmit AS, SIM phase shifts, and PA, to achieve EM domain-based frame-level IEP, as described in Algorithm~\ref{alg:3}.

\section{Numerical Results}\label{Sec.V}
Simulations are performed to evaluate the performance of the proposed SIM-enabled IEP scheme, which incorporates the ROM algorithm for optimizing SIM phase shifts and the ROM-AO algorithm for jointly optimizing the transmit AS, SIM phase shifts, and PA. The evaluation is conducted in terms of the SER, the objective function, and the receiver constellation diagrams.

\subsection{Simulation Setup}\label{Sec.V.1}
\begin{figure}[htbp]
	\centering{}\includegraphics[width=3.5in]{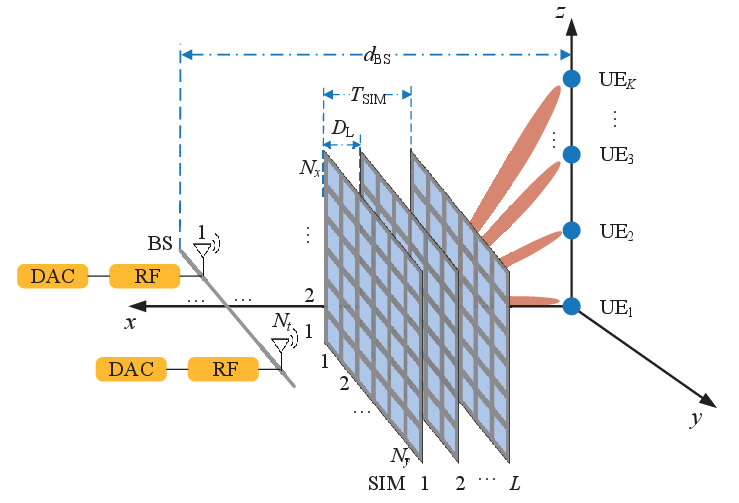}
	\caption{Simulation setup of SIM-empowered MU-MISO communication system.}
	\label{fig:3}
\end{figure}

As shown in Fig. \ref{fig:3}, a SIM-assisted MU-MISO downlink communication system is considered. The BS is deployed parallel to the $y$-axis at a distance of $d_{\mathrm{BS}}=10\ \mathrm{m}$, and its center is located on the $x$-axis, transmitting signals at frequency $f_{0}$. There is a $L$-layer SIM centered around the $x$-axis, and is deployed in the $y$-$z$ plane. The $K$ users equipped with $N_r$ antennas are uniformly distributed on the $z$-axis with spacing $d_{\mathrm{UE}}=10$ m. The macroscopic large-scale fading is modeled as $\rho_k=C_{\mathrm{0}}d_k^{-\alpha}$, where $C_{\mathrm{0}}$ is the free space path loss at the reference distance of 1 m and $\alpha$ is the path loss exponent. 
The main parameters are presented in Table \ref{table1}.

\begin{table}[htbp]
    \large
    \renewcommand{\arraystretch}{1.5}
	\centering
	\caption{Simulation Parameters}
    \begin{adjustbox}{width=0.45\textwidth}
    \begin{tabular}{|m{4cm}<{\centering}|m{2cm}<{\centering}|m{4cm}<{\centering}|m{2cm}<{\centering}|} \hline
		Parameter & Value & Parameter & Value\\ \hline\hline
		$f_{0}$ & 30 GHz &	$N_r$ & 1  \\ \hline
		$\Delta$ & $\lambda/2$ & $d_{\mathrm{UE}}$ & 10 m  \\ \hline
		$T_{\mathrm{SIM}}$ & 10 $\lambda$  &
	  $C_{\mathrm{0}}$ & $-20\ \mathrm{dB}$ \\ \hline
		Modulation mode & QPSK &
		$\alpha$ & 3.5  \\ \hline
		$\epsilon$ & $10^{-3}$  &
		$U$ & 128 \\ \hline
        $\alpha_a$ & 1.6623~\cite{1094911} & $\beta_a$ & 0.0552~\cite{1094911} \\ \hline
        $\alpha_{\phi}$ & 0.1533~\cite{1094911} & $\beta_{\phi}$ & 0.3456~\cite{1094911} \\ \hline
        $\varepsilon$ & $10^{-1}$ & $\varepsilon_p$ & $10^{-1.5}$ \\ \hline
	\end{tabular}
    \end{adjustbox}
	\label{table1}
\end{table}

\subsection{Simulation Results for ROM Algorithm}\label{Sec.V.2}
\begin{figure}[htbp]
	\centering
	\subfigure[]
	{\includegraphics[width=1.8in]{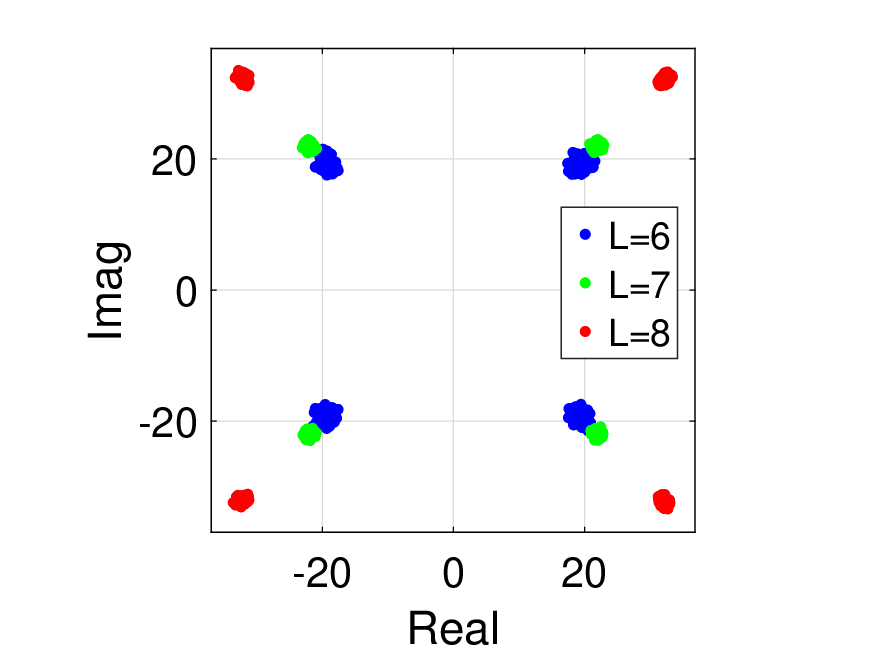}}
    \hspace{-0.8cm}
	\subfigure[]
	{\includegraphics[width=1.8in]{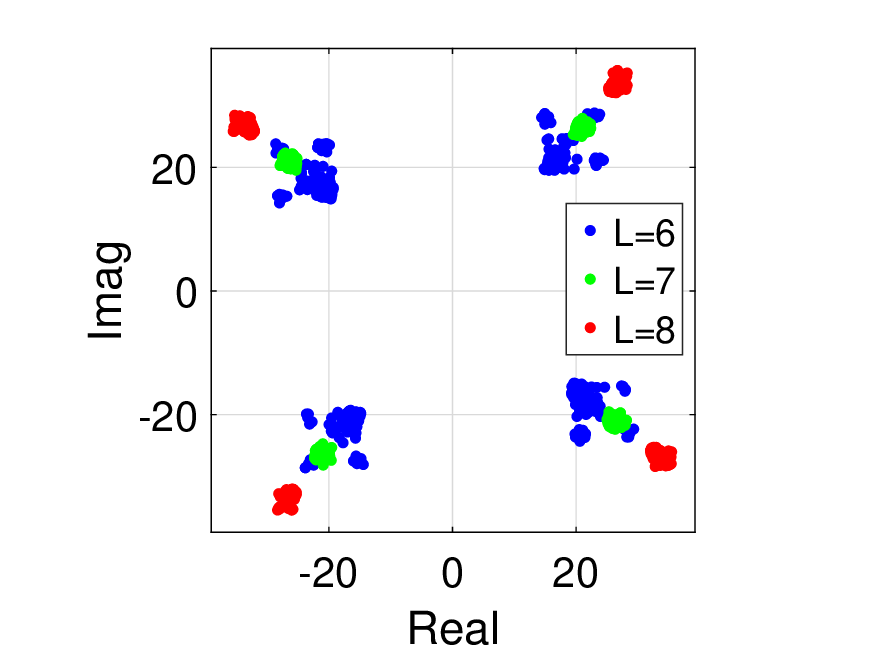}}
	\caption{Constellation diagrams for different SIM layers under NLD-aware and NLD-unaware schemes, with $K = 5$, $N = 36$: (a) NLD-aware; (b) NLD-unaware.}
	\label{fig:4}
\end{figure}

\begin{figure}[ht]
	\centering{}\includegraphics[width=3.45in]{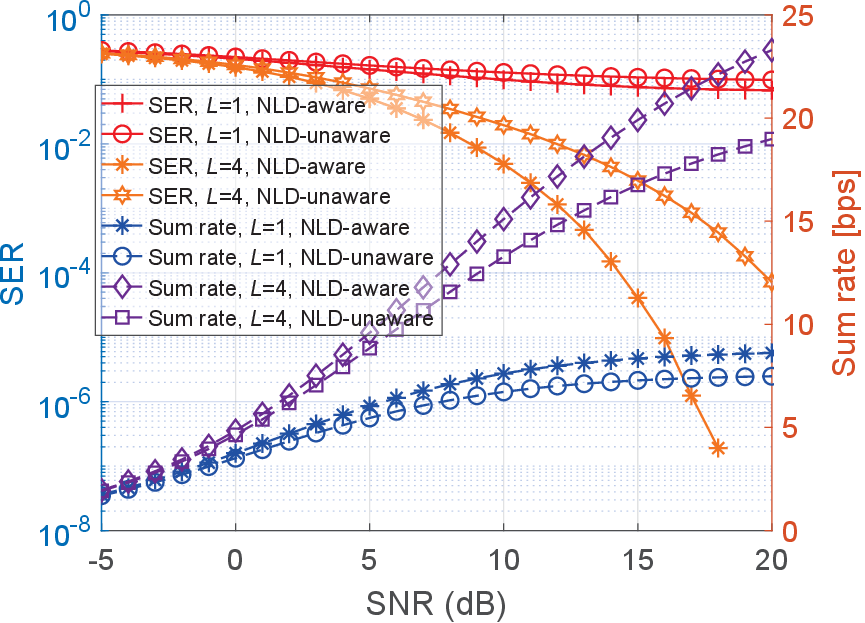}
	\caption{SER and sum rate vs. SNR, with $K = 5$, $N = 36$.}
	\label{fig:5}
\end{figure}

Fig.~\ref{fig:4} illustrates the constellation points for NLD-aware and NLD-unaware schemes. Fig.~\ref{fig:4}(a) plots the distribution of constellation points generated by different values of $L$ in the MU-MISO communication system with $K = 5$ and $N = 36$ under the NLD-aware scheme. It is seen that the resulting constellation points are closely concentrated around the ideal constellation.
The constellation diagram generated by the NLD-unaware scheme is shown in
Fig.~\ref{fig:4}(b), where the angular shifts in the constellation points are apparent. Compared to the NLD-unaware IEP, the NLD-aware IEP leads to a larger safety margin, which indicates that the NLD-aware optimization of the SIM phase shifts effectively compensates for the NLD effect caused by the power amplifiers, eliminating the need for traditional digital pre-distortion~\cite{bhargava2023pre,10258313}. Meanwhile, Fig.~\ref{fig:4} also proves that more SIM layers provide higher DoF to effectively utilize MUI and push the constellation points away from decision boundaries into the CI region. \par

Fig.~\ref{fig:5} depicts the SER and the sum rate performance vs. SNR for both single-layer RIS, i.e., $L = 1$ and multi-layer SIM with $L = 4$, under both NLD-aware and NLD-unaware IEP schemes, in the considered MU-MISO communication system with $K = 5$ and $N = 36$. The results demonstrate the fundamental limitations of the single-layer structure. Specifically, the SER curves for $L = 1$ exhibit a pronounced error floor at high SNRs, and the sum rate also saturates early. This can be attributed to its limited DoFs and physical constraints, which prevent effective spatial processing capability and interference exploitation in multi-user scenarios. In contrast, the proposed SIM can provide substantial wave-based processing capabilities, significantly enhancing communication reliability and spatial multiplexing. Notably, when the SNR reaches 10 dB, the NLD-aware SIM achieves a 13.07 dB SER performance gain and approximately doubles the sum rate, compared to the NLD-aware RIS. \par

Moreover, the proposed NLD-aware scheme consistently outperforms its NLD-unaware counterpart across all tested scenarios. 
The NLD-aware SIM design provides an SNR advantage of nearly 4 dB and 2 dB at an SER of $10^{-3}$ and a sum rate of 15 bps, respectively, compared to its NLD-unaware counterpart. 
Even for the single-layer RIS, which exhibits an SER error floor and a sum rate saturation, the NLD-aware design achieves significantly performance improvements over NLD-unaware version, with a 45.00\% reduction in SER at the error floor region and a 15.06\% enhancement in sum rate at the saturation stage.
The consistent performance gain observed in the NLD-aware scheme demonstrates the SIM's capability to compensate for power amplifier nonlinearities. This compensation is particularly critical for the proposed IEP framework, as it directly preserves the precise phase relationships required for constructive interference, thereby maintaining a large safety margin and enabling the full performance potential of the system. \par

\begin{figure}[ht]
	\centering{}\includegraphics[width=3.5in]{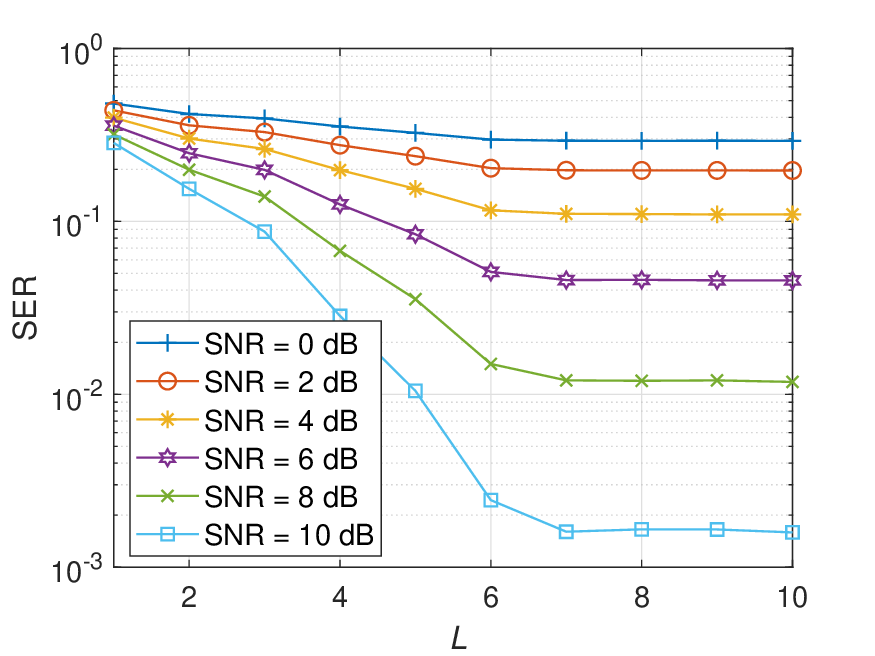}
	\caption{SER vs. number of layers in different SNR, with $K = 5$, $N = 36$.}
	\label{fig:6}
\end{figure}

Fig.~\ref{fig:6} studies the impact of the number of SIM layers on the SER. 
An increase in the number of layers leads to significant improvement in SER performance, which demonstrates the wave-domain signal processing capability of multi-layer architecture.
Compared to $L = 1$, the 6-layer SIM demonstrates 8.46 dB and 20.66 dB SER performance gain, when SNR reaches 6 dB and 10 dB, respectively. This trend shows that at lower SNRs, the performance is noise-limited, hence the gain from sophisticated MUI exploitation and precise beamforming is constrained by the additive noise. In contrast, at higher SNRs, the superior interference management capabilities of the multi-layer SIM fully unleashed.
As $L$ increases to 7, the SER performance reaches a saturation point, showing no significant improvement over $L = 7$ configuration. This saturation indicates that further hardware resources could be more effectively allocated to other system parameters. \par

\begin{figure}[ht]
	\centering{}\includegraphics[width=3.5in]{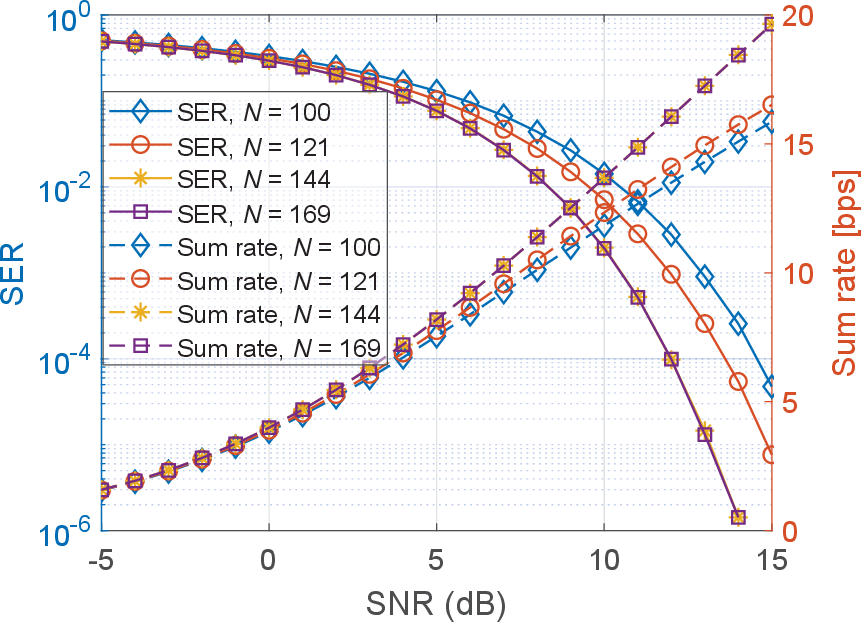}
	\caption{SER and sum rate comparison across different values of $N$, with $K = 4$, $L = 3$.}
	\label{fig:7}
\end{figure}

Fig.~\ref{fig:7} presents a detailed comparison of SER and sum rate performance across different number of meta-atoms $N$ per SIM layer, under configurations of $K = 4$ and $L = 3$. We investigate the roles of $N$ as a crucial source of effective DoFs for providing spatial processing capability in multi-user communications, particularly when the number of SIM layers is limited, e.g., $L = 3$. It is observed that as $N$ increases from 100 to 121 and then to 144, both the SER and sum rate performance improve significantly. The SER performance improves by 6.73 dB and 22.52 dB, respectively, while the sum rate improves by 4.36\% and 22.30\%, respectively. This is attributed to the enhanced DoFs, which enables more effective exploitation of MUI through superior wave-based beamforming.
As $N$ is further increased from 144 to 169, the SER and sum rate curves nearly overlap, indicating a performance saturation. This trend highlights a critical insight into practical design implications, e.g., the performance saturates as $N$ increases beyond a certain point, and suggests an optimal range for $N$ under practical hardware constraints. \par

\begin{figure}[t]
	\centering{}\includegraphics[width=3.5in]{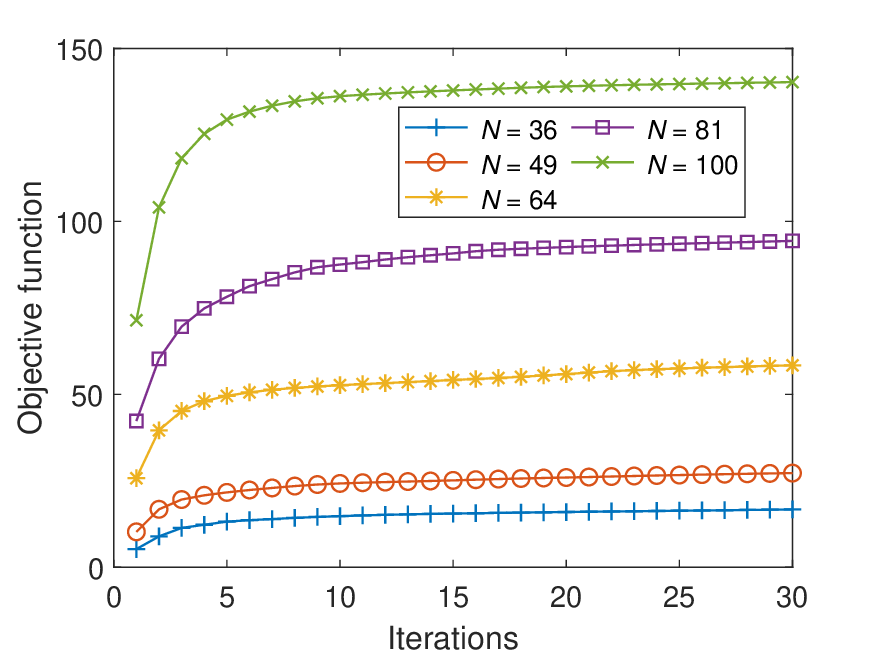}
	\caption{Objective function vs. the number of iterations under different number of meta-atoms per layer, with $K = 4$, $L = 4$.}
	\label{fig:8}
\end{figure}

Fig.~\ref{fig:8} shows the convergence behavior of the proposed ROM algorithm.
First, the curves confirm the algorithm's robustness across all configurations, exhibiting monotonic and stable convergence.
As the number of meta-atoms $N$ increases from 36 to 100, the relative change at convergence monotonically decreases from 0.45\% to 0.07\%, while the number of iterations required to achieve 98\% of the final performance reduces from 26 to 14. This reflects the more favorable optimization geometry of high-dimensional oblique manifolds, where the expanded parameter space enables smoother gradient paths and more efficient convergence toward high-quality solutions, thus effectively mitigating the curse of dimensionality and circumventing low-dimensional local plateaus. These characteristics collectively confirm the proposed ROM algorithm's robustness, scalability, and convergence efficiency in leveraging hardware resources.
Furthermore, the converged objective value, i.e., the minimum safety margin, increases significantly with $N$, i.e., about 8 times as $N$ increases from 36 to 100. This demonstrates that a larger $N$ enhances the available DoFs, enabling the SIM to more precisely exploit the MUI for pushing received symbols farther away from decision boundaries, thereby achieving the greater safety margin. \par

\begin{figure}[t]
    \centering
    \begin{minipage}[t]{1.0\linewidth}
    \centering
        \begin{tabular}{@{\extracolsep{\fill}}c@{}c@{}@{\extracolsep{\fill}}}
            \includegraphics[width=0.5\linewidth]{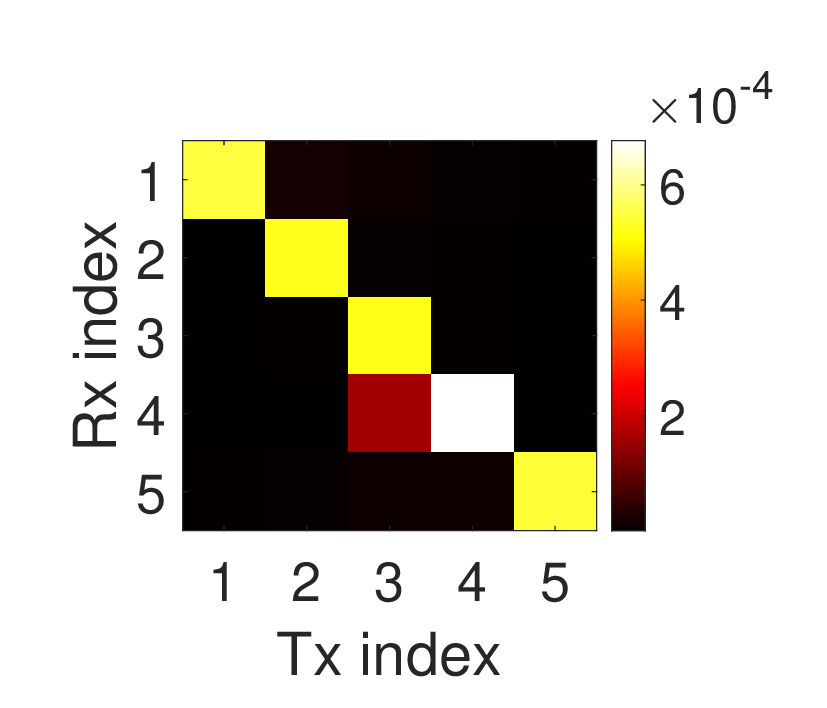} &
            \includegraphics[width=0.5\linewidth]{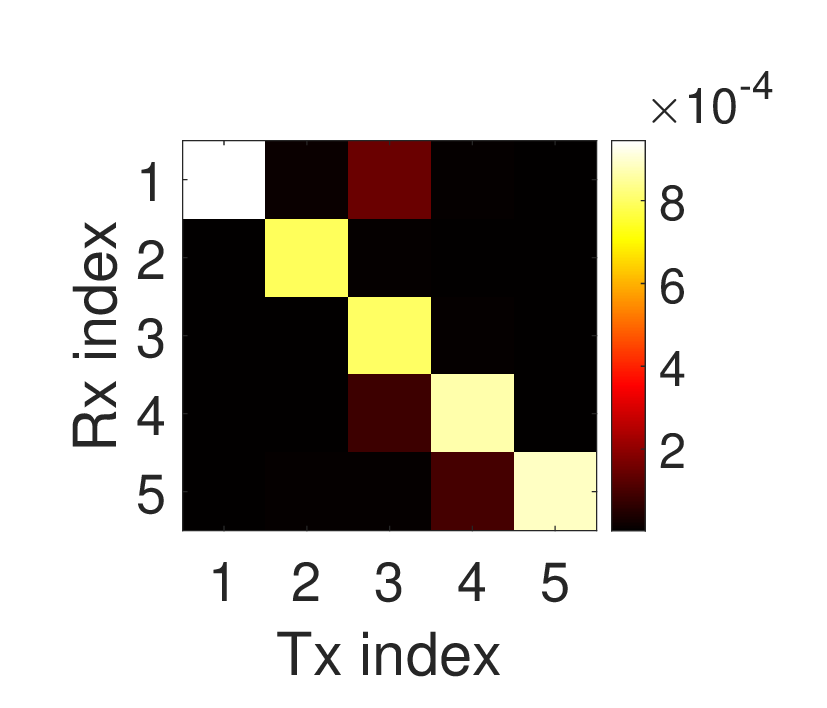}\\
            (a) $N=81$ & (b) $N=100$\\
        \end{tabular}
    \end{minipage}
    \begin{minipage}[t]{1.0\linewidth}
    \centering
        \begin{tabular}{@{\extracolsep{\fill}}c@{}c@{}@{\extracolsep{\fill}}}
            \includegraphics[width=0.5\linewidth]{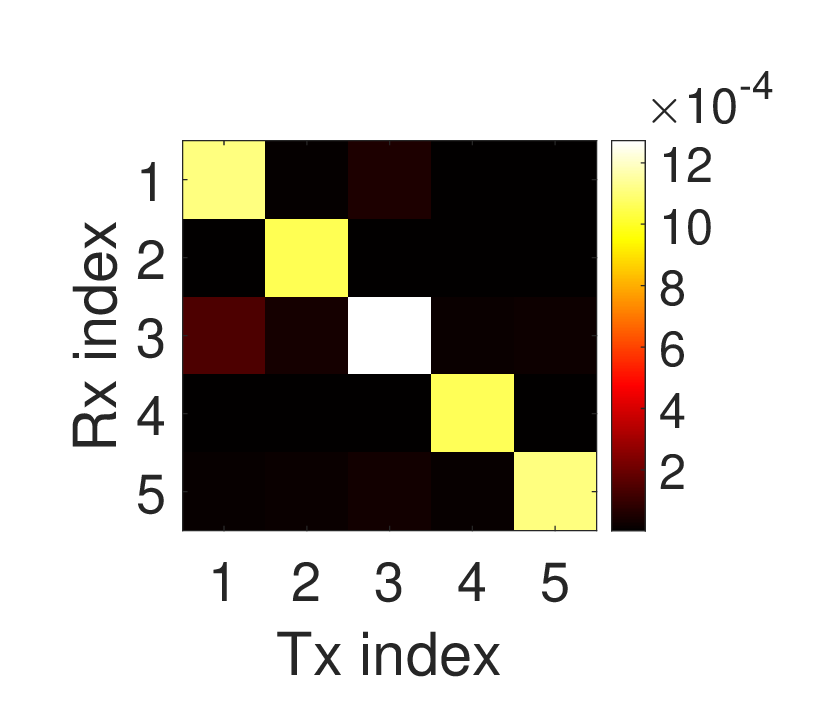} &
            \includegraphics[width=0.5\linewidth]{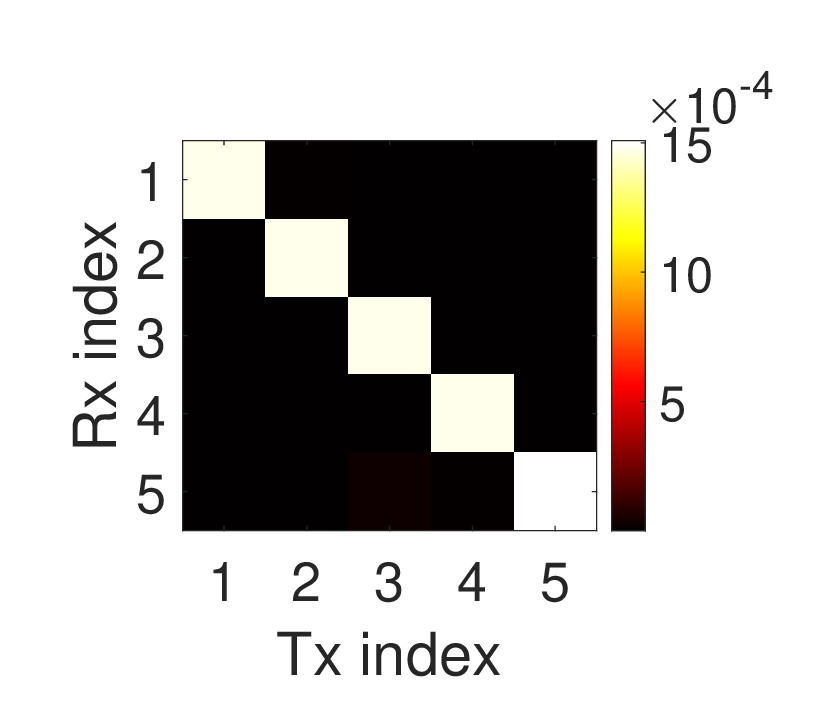}\\
            (c) $N=121$ & (d) $N=144$\\
        \end{tabular}
    \end{minipage}
    \caption{Channel gain heat maps for different values of $N$, with $K = 5$, $L = 4$.}
    \label{fig:9}
\end{figure}

Fig.~\ref{fig:9} shows the channel gain heat maps for different values of $N$.
It is observed that with $N$ increasing, the diagonal gains of the channel matrix gradually increase, while the off-diagonal interference components gradually decrease.
Specifically, with $N$ increasing from 81 to 144, the difference strength between average diagonal and non-diagonal elements in the channel matrix increases from 16.07 dB to 24.24 dB. This demonstrates that the increased $N$ can enhance the SIM's wave-based beamforming capability, not only forming more focused and precise beams, but also more effectively sculpting the MUI into a constructive force.
Furthermore, the variance of the diagonal elements decreases by 12.83 dB as $N$ increases from 81 to 144. This indicates that the proposed IEP strategy can offer a novel pathway towards achieving greater fairness in multi-user communication scenarios by controlling beamforming and leveraging the traditionally detrimental MUI to help users with poorer channel conditions, instead of the selfish resource allocation in traditional interference suppression methods. \par

\begin{figure}[t]
	\centering{}\includegraphics[width=3.5in]{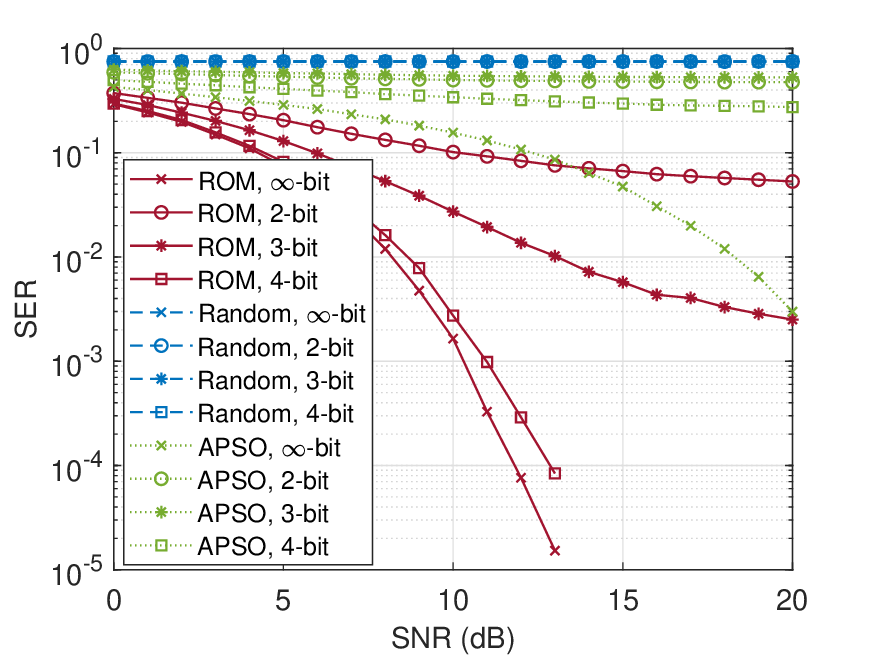}
	\caption{SER comparison of different SIM phase shift optimization methods.}
	\label{fig:10}
\end{figure}
 
Furthermore, Fig. \ref{fig:10} compares the algorithm performance with two benchmark schemes, including the adaptive particle swarm optimization (APSO) algorithm~\cite{4812104} and the random SIM phase shifts, in the MU-MISO system considering four users. A 5-layer SIM with $9\times9$ meta-atoms arranged in uniform planar array (UPA) per layer is employed. 
The SER performance of our proposed scheme dramatically outperforms the benchmark schemes. The worst performance is observed with the random SIM phase shifts. For the APSO algorithm, the high dimensionality of the search space increases the probability of premature convergence to local optima. This prevents the APSO from finding globally optimal SIM phase shifts. For the random SIM phase shifts without any optimization, it fails to utilize the MUI, leading to poor phase alignment.
On the other hand, the optimal discrete SIM phase shifts for each scheme are determined by selecting the nearest discrete phase values to the optimal continuous SIM phase shifts. Moreover, the SER results for 2-bit and 3-bit resolutions are significantly worse than that of the continuous phase shifts.
The discrete phase shifts of 4-bit can relatively approach the SER performance achieved by the continuous phase shifts. \par

\begin{figure}[t]
	\centering{}\includegraphics[width=3.5in]{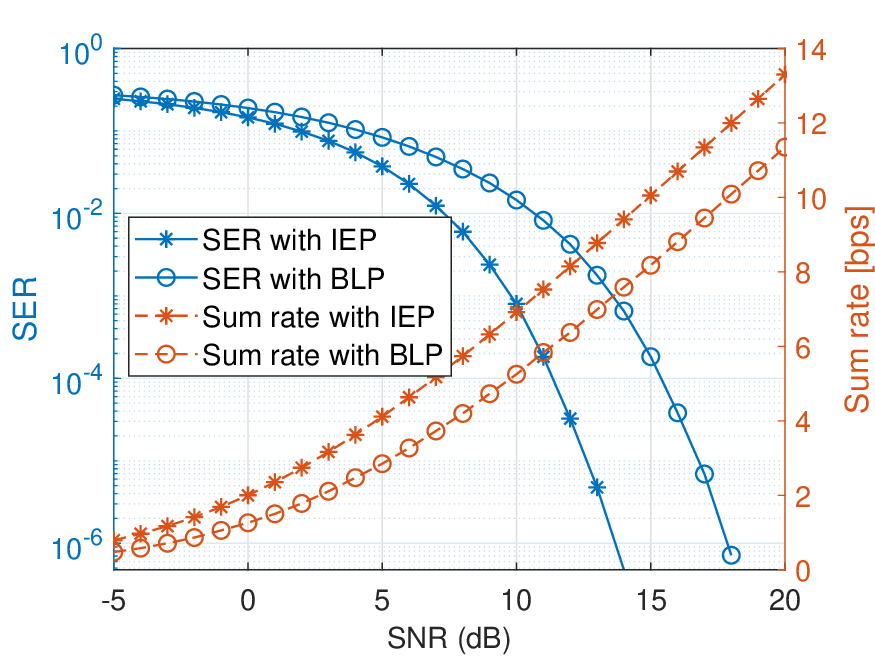}
	\caption{SER and sum rate comparison for IEP and BLP schemes, with $K = 2$, $L = 6$, $N = 81$.}
	\label{fig:11}
\end{figure}

Fig.~\ref{fig:11} compares SER and sum rate between the proposed IEP scheme and traditional BLP, e.g., zero-forcing (ZF) precoder, within an MU-MISO communication system considering two users.
The proposed IEP scheme demonstrates consistently superior SER and sum rate performance compared to traditional BLP technology.
At the SNR of 10 dB, it provides SER and sum rate performance gains of 12.60 dB and 1.20 dB, respectively, over the traditional ZF precoder. In contrast, to achieve a benchmark SER of $10^{-3}$ and a sum rate of 10 bps, the ZF precoder exhibits an SNR penalty of approximately 4 dB and 3 dB, respectively. The superiority of our method arises because the IEP constructively utilizes MUI to push the received symbols farther away from the decision boundaries, thereby enhancing the detection reliability and system robustness. By contrast, ZF precoding simply nulls the MUI. \par

\subsection{Simulation Results for ROM-AO Algorithm}\label{Sec.V.3}
\begin{figure}[t]
    \centering
    \begin{minipage}[t]{1.0\linewidth}
    \centering
        \begin{tabular}{@{\extracolsep{\fill}}c@{}c@{}@{\extracolsep{\fill}}}
            \includegraphics[width=0.5\linewidth]{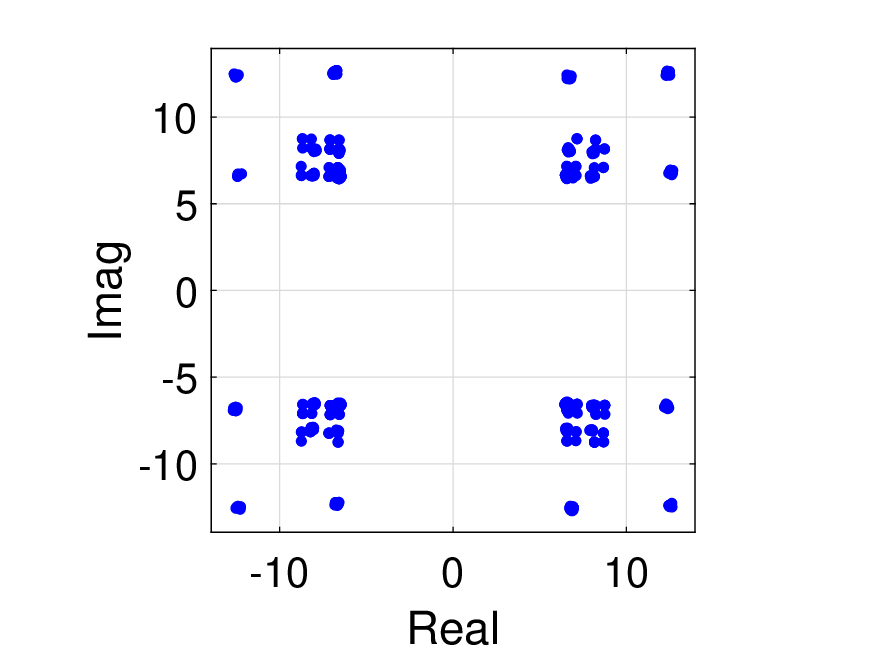} &
            \includegraphics[width=0.5\linewidth]{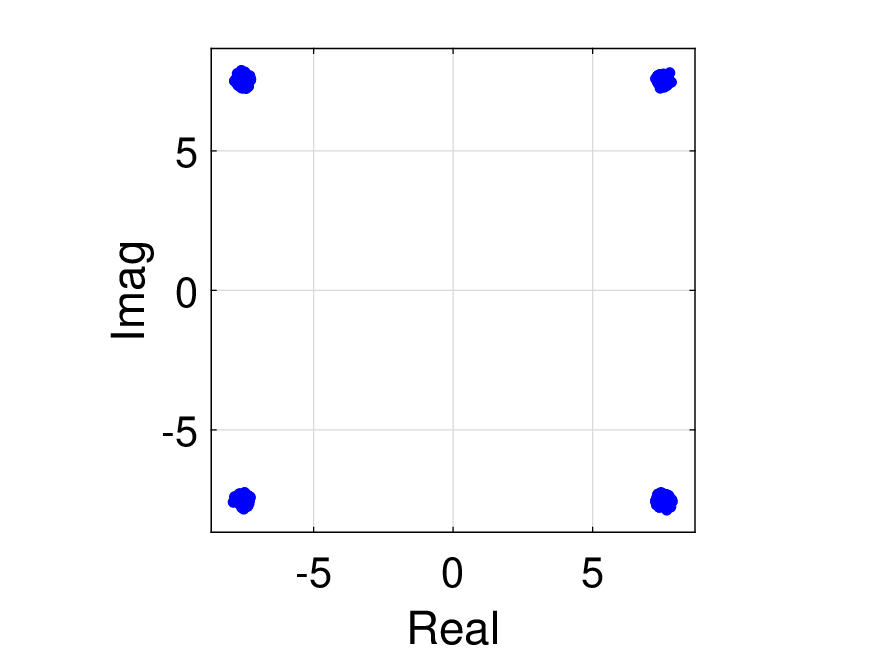}\\
            (a) w/ AS - w/o PA & (b)w/ AS - w/ PA\\
        \end{tabular}
    \end{minipage}
    \begin{minipage}[t]{1.0\linewidth}
    \centering
        \begin{tabular}{@{\extracolsep{\fill}}c@{}c@{}@{\extracolsep{\fill}}}
            \includegraphics[width=0.5\linewidth]{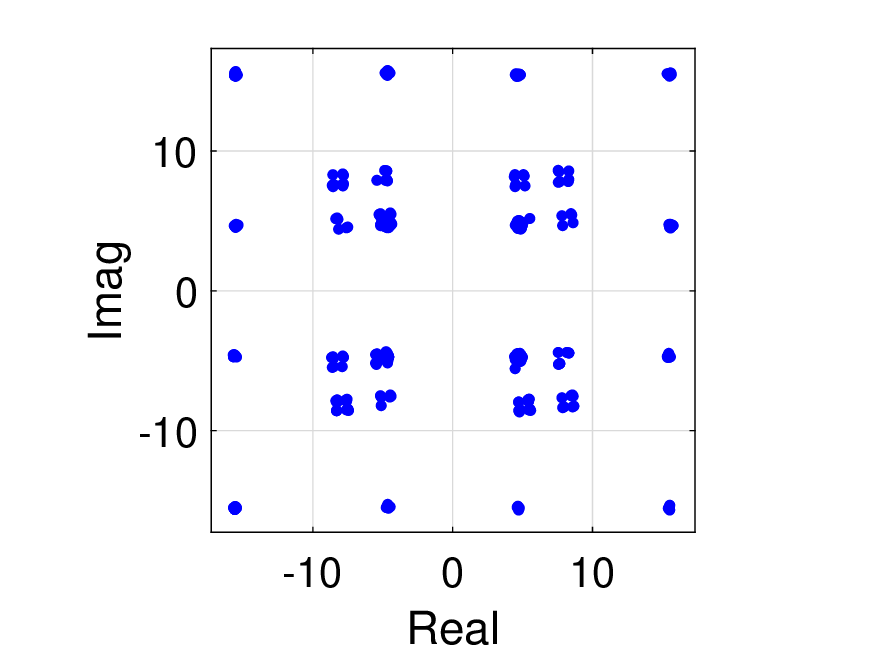} &
            \includegraphics[width=0.5\linewidth]{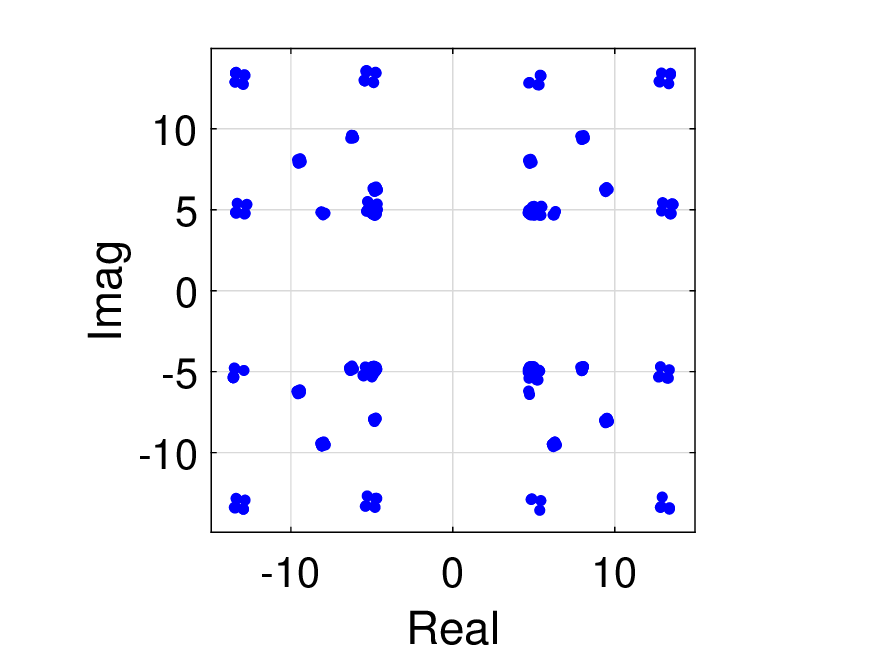}\\
            (c) w/o AS - w/o PA & (d) w/o AS - w/ PA\\
        \end{tabular}
    \end{minipage}
    \caption{Constellation diagrams for different AS and PA strategies, with $K = 4$, $L = 4$, $N = 36$.}
    \label{fig:12}
\end{figure}
 
Fig.~\ref{fig:12} shows the constellation diagrams for different AS and PA strategies, and the four strategies represent the performance of:
\begin{enumerate}
    \item greedy safety margin-based AS algorithm without PA;
    \item greedy safety margin-based AS algorithm with PA;
    \item selecting antennas in sequence without PA;
    \item selecting antennas in sequence with PA.
\end{enumerate}
From Fig.~\ref{fig:12}(c), it is concluded that the AS and PA strategies can provide higher DoF to adapt to dynamic channel conditions. The resulting constellation points are more stably distributed within the CI region. 
Comparing Figs.~\ref{fig:12}(a) and (d), we observe that the AS strategy exerts a more substantial influence on IEP performance than PA, with 47.54\% and 6.67\% improvements of the minimum safety margin, respectively, compared to the case with no strategy. \par
  
\begin{figure}[t]
	\centering{}\includegraphics[width=3.5in]{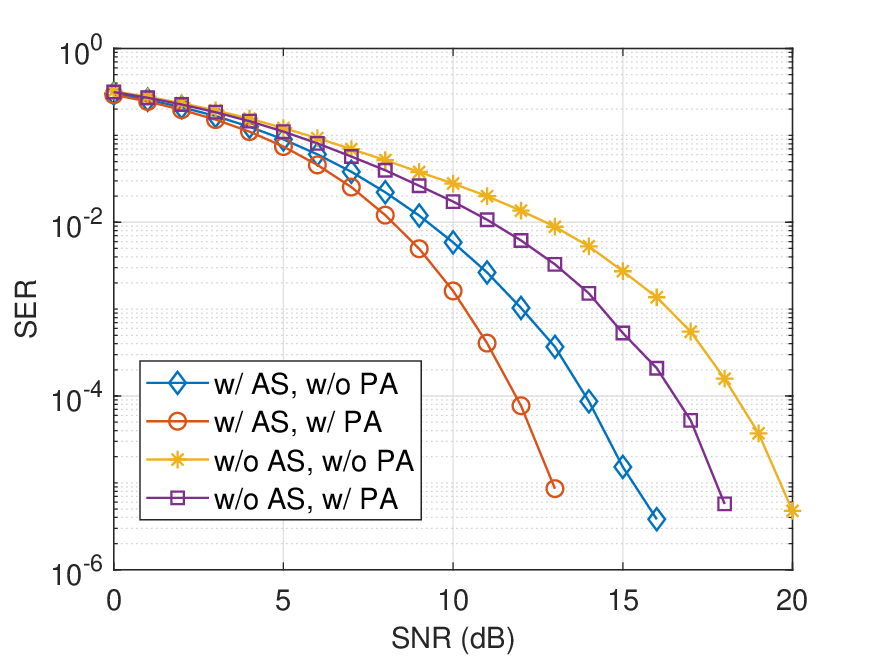}
	\caption{SER vs. SNR for different AS and PA strategies, with $K = 4$, $L = 4$, $N = 36$.}
	\label{fig:13}
\end{figure}

Fig.~\ref{fig:13} plots the SER vs. SNR for the aforementioned four AS and PA strategies. It is seen that the SER achieved by the greedy safety margin-based AS algorithm consistently outperforms that achieved by the sequential selection, regardless of whether PA is applied. Meanwhile, PA can lead to better SER performance compared to schemes without PA.
In particular, when the SNR reaches 12 dB, the strategy with AS and PA demonstrates a 20 dB improvement compared to the case without any strategy. Furthermore, it is observed that the strategy with AS achieves an approximate 11 dB improvement over the case without any strategy, while the strategy with PA offers an approximation of 3 dB improvement under the same conditions. Therefore, compared to PA, AS has a more significant impact on enhancing SER performance. \par

\section{Conclusion}\label{Sec.VI}
In this paper, we designed a SIM-enabled IEP architecture for MU-MISO systems. By leveraging the wave-based analog computing capacity of the programmable SIM, the MUI can be converted to the CI, and the NLD caused by the power amplifiers on the transmitted signals can be effectively compensated. 
We formulated an EM domain-based frame-level IEP problem to maximize the minimum safety margin by optimizing SIM phase shifts per frame. To efficiently update SIM phase shifts, an ROM algorithm was proposed. Subsequently, we investigated an AS scenario in the SIM-enabled IEP system. With the AS, flexible DoF was achieved by the SIM for dynamic wireless environments. This yielded a joint optimization of transmit AS, SIM phase shift design, and PA. To solve this problem, an ROM-AO algorithm was proposed, where a greedy safety margin-based AS algorithm was designed to select the antennas. 
Simulations verified our algorithms. Moreover, the SIM can effectively compensate for the NLD effect and achieve superior performance compared with anti-interference benchmarks.


\bibliography{ref.bib}

\bibliographystyle{IEEEtran}
\begin{IEEEbiography}[{\includegraphics[width=1in,height=1.25in,clip,keepaspectratio]{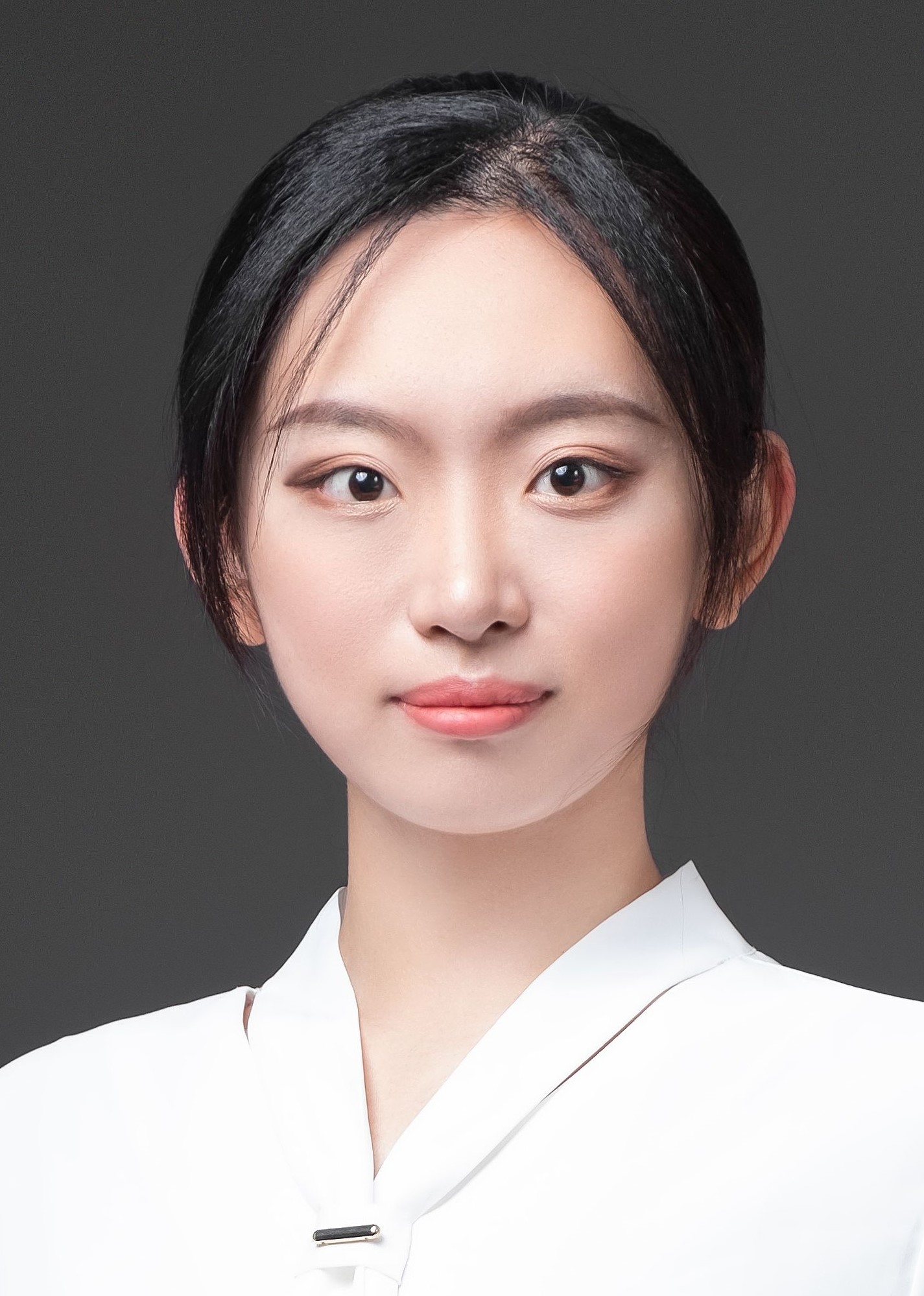}}]{Hetong Wang}
received the B.S. and M.S. degrees from the School of Telecommunications Engineering, Xidian University, Xi'an, China, in 2020 and 2023, respectively. She is currently pursuing the Ph.D. degree at the School of Information and Communication Engineering, Beijing University of Posts and Telecommunications (BUPT), Beijing, China. Her current research interests include physical layer security, reconfigurable intelligent surface, stacked intelligent metasurface, and machine learning.
\end{IEEEbiography}

\begin{IEEEbiography}[{\includegraphics[width=1in,height=1.25in,clip,keepaspectratio]{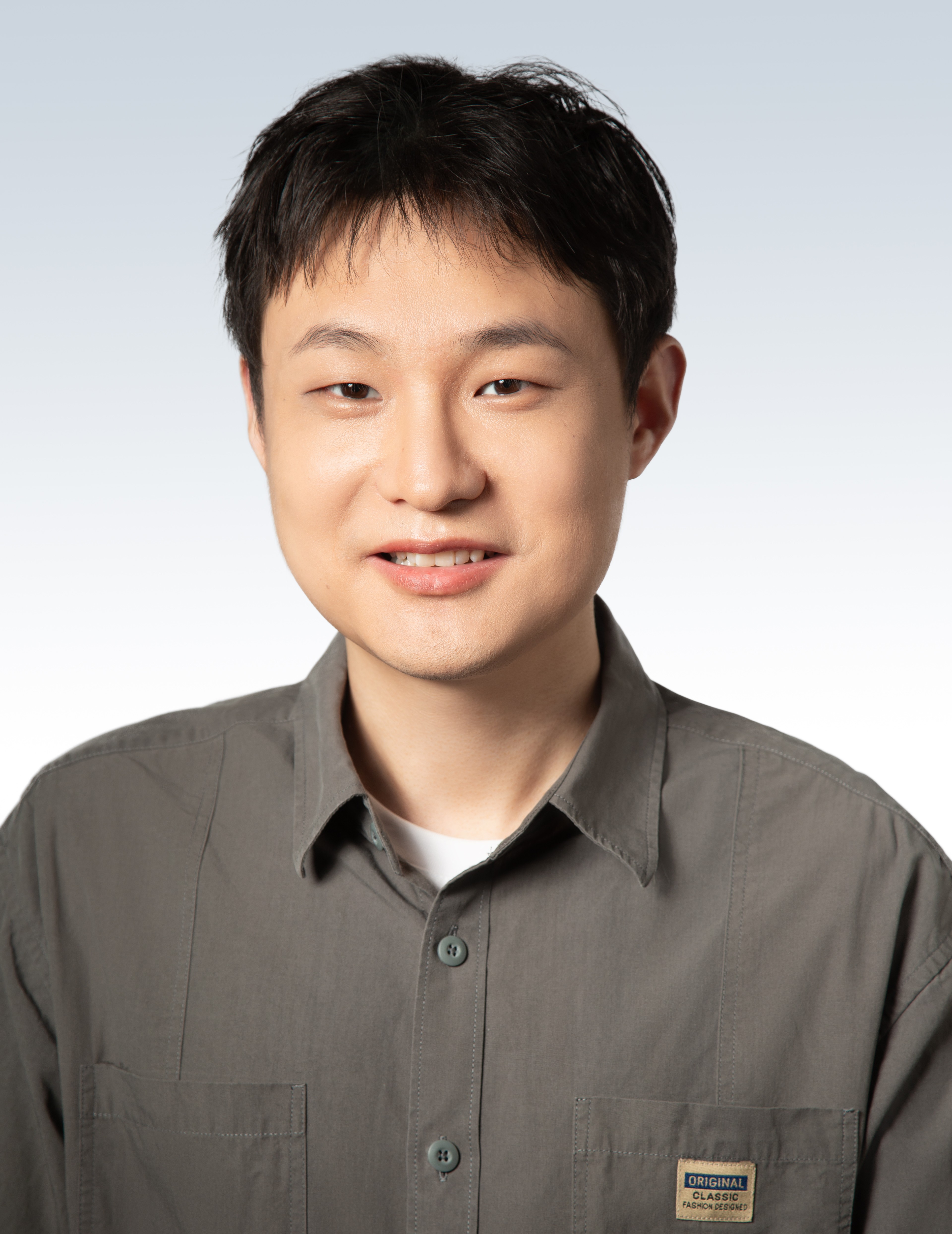}}]{Yashuai Cao}
received the B.E. and Ph.D. degrees in communication engineering from Chongqing University of Posts and Telecommunications (CQUPT) and Beijing University of Posts and Telecommunications (BUPT), China, in 2017 and 2022, respectively. From 2022 to 2023, he was a lecturer in the Department of Electronics and Communication Engineering, North China Electric Power University (NCEPU), Baoding. From 2023 to 2025, he was a Postdoctoral Research Fellow with the Department of Electronic Engineering, Tsinghua University, Beijing, China. He is currently a Distinguished Associate Professor with the School of Intelligence Science and Technology, University of Science and Technology Beijing, Beijing, China. His research interests include Intelligent Reflecting Surface, Stacked Intelligent Metasurface, Environment-Aware Communications, and Channel Twinning.
\end{IEEEbiography}

\begin{IEEEbiography}[{\includegraphics[width=1in,height=1.25in,clip,keepaspectratio]{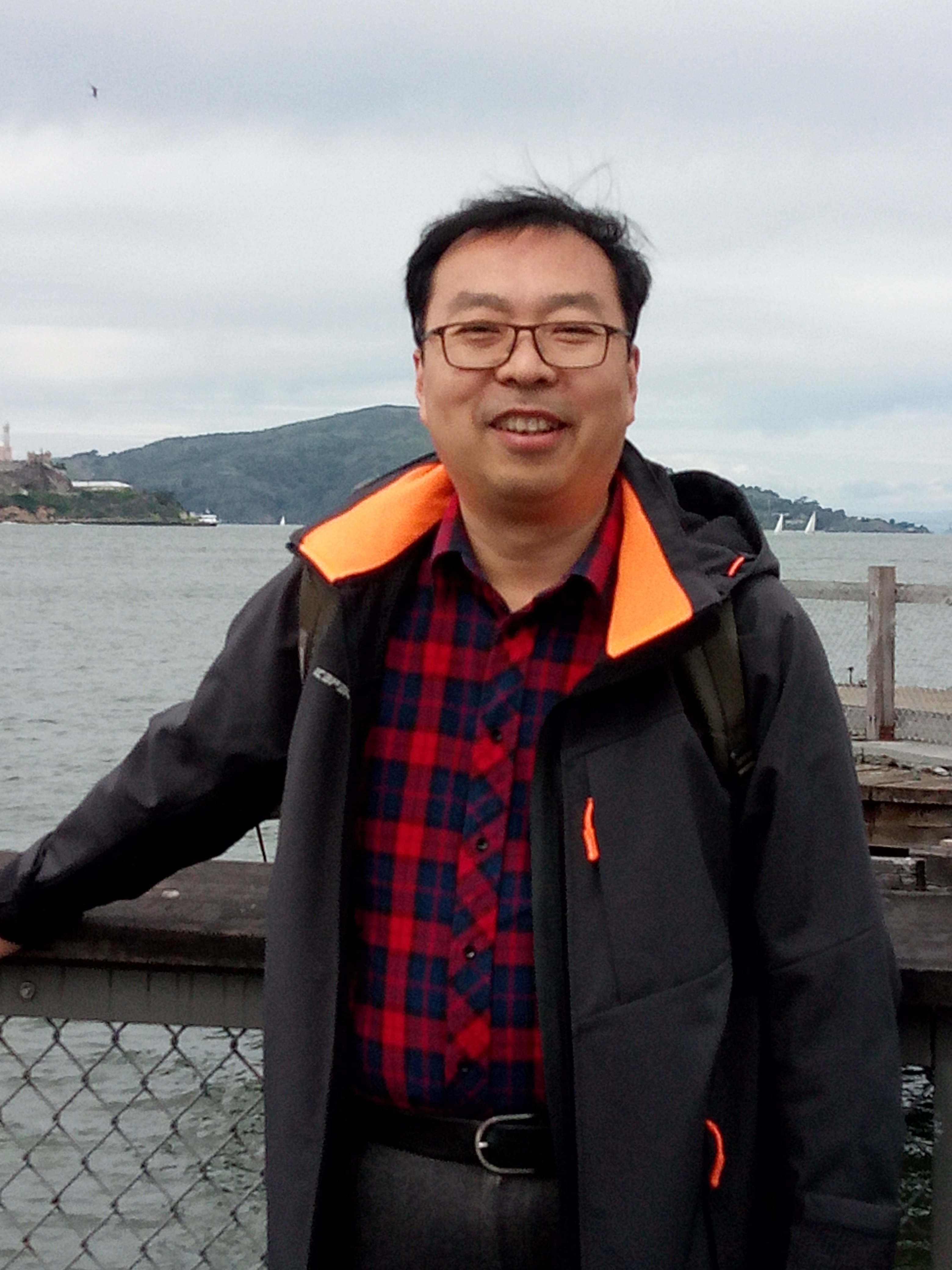}}]{Tiejun Lv}
received the M.S. and Ph.D. degrees in electronic engineering from the University of Electronic Science and Technology of China (UESTC), Chengdu, China, in 1997 and 2000, respectively. From January 2001 to January 2003, he was a Postdoctoral Fellow at Tsinghua University, Beijing, China. In 2005, he was promoted to Full Professor at the School of Information and Communication Engineering, Beijing University of Posts and Telecommunications (BUPT). From September 2008 to March 2009, he was a Visiting Professor with the Department of Electrical Engineering at Stanford University, Stanford, CA, USA. He is the author of four books, one book chapter, more than 160 published journal papers and 200 conference papers on the physical layer of wireless mobile communications. His current research interests include signal processing, communications theory and networking. He was the recipient of the Program for New Century Excellent Talents in University Award from the Ministry of Education, China, in 2006. He received the Nature Science Award from the Ministry of Education of China for the hierarchical cooperative communication theory and technologies in 2015 and the Shaanxi Higher Education Institutions Outstanding Scientific Research Achievement Award in 2025.
\end{IEEEbiography}

\begin{IEEEbiography}[{\includegraphics[width=1in,height=1.25in,clip,keepaspectratio]{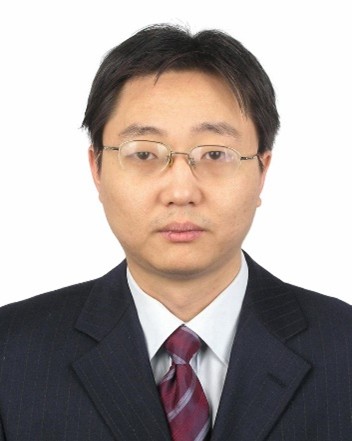}}]{Jintao Wang}
(Fellow, IEEE) received the B.Eng. and Ph.D. degrees in electrical engineering both from Tsinghua University, Beijing, China, in 2001 and 2006, respectively. From 2006 to 2009, he was an Assistant Professor in the Department of Electronic Engineering at Tsinghua University. Since 2009, he has been an Associate Professor and Ph.D. Supervisor. He is the Standard Committee Member for the Chinese national digital terrestrial television broadcasting standard. His current research interests include space-time coding, MIMO, and OFDM systems. He has published more than 100 journal and conference papers and holds more than 40 national invention patents.
\end{IEEEbiography}

\begin{IEEEbiography}[{\includegraphics[width=1in,height=1.25in,clip,keepaspectratio]{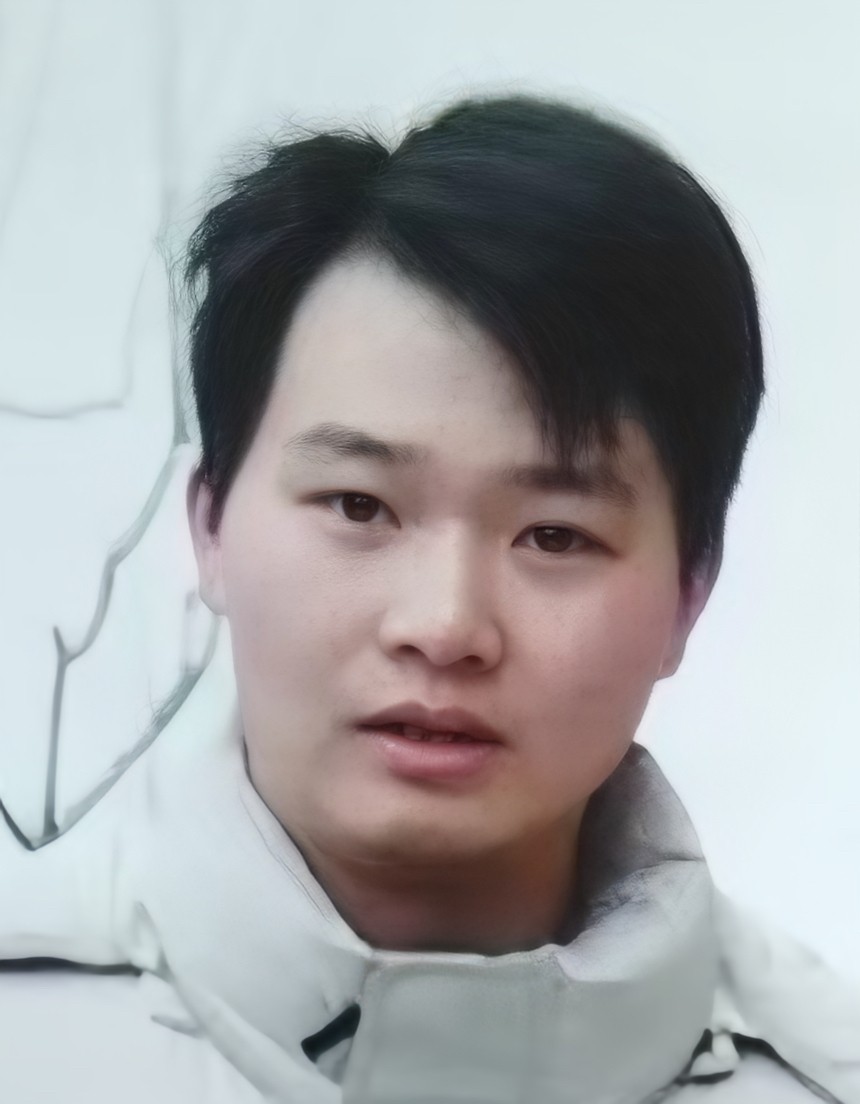}}]{Wei Ni}
received his Ph.D. degree in Communication Engineering from Shanghai Jiao Tong University, Shanghai, China, in 2007. He is a Senior Research Engineer at the School of Information Science and Technology, Fudan University, Shanghai, China. His research interests encompass a wide range of wireless communication research and engineering areas, including channel measurement and modeling, digital signal processing, antenna array processing, embedded programming, localization and synchronization, remote sensing, cellular systems, network planning and optimization, Internet of Things, and applied machine learning.
\end{IEEEbiography}

\begin{IEEEbiography}[{\includegraphics[width=1in,height=1.25in,clip,keepaspectratio]{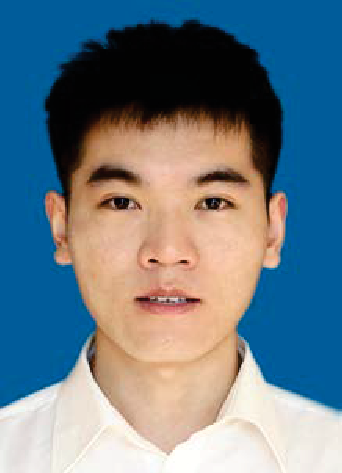}}]{Jiancheng An}
(Senior Member, IEEE) received the B.S. degree in Electronics and Information Engineering and the Ph.D. degree in Information and Communication Engineering from the University of Electronic Science and Technology of China (UESTC), Chengdu, China, in 2016 and 2021, respectively. From 2019 to 2020, he was a Visiting Scholar with the Next-Generation Wireless Group, University of Southampton, U.K. From 2021 to 2023, he was a Post-Doctoral Research Fellow with the Engineering Product Development (EPD) Pillar, Singapore University of Technology and Design (SUTD), Singapore. He is currently a Research Fellow with the School of Electrical and Electronics Engineering, Nanyang Technological University (NTU), Singapore. Dr. An received the IEEE International Conference on Communications (ICC) 2023 Best Paper Award. He is the co-inventor of six patents and has published over 100 research papers in peer-reviewed international journals and conferences. His research interests include stacked intelligent metasurfaces (SIM), flexible intelligent metasurfaces (FIM), and wave-based computing.
\end{IEEEbiography}

\begin{IEEEbiography}[{\includegraphics[width=1in,height=1.25in,clip,keepaspectratio]{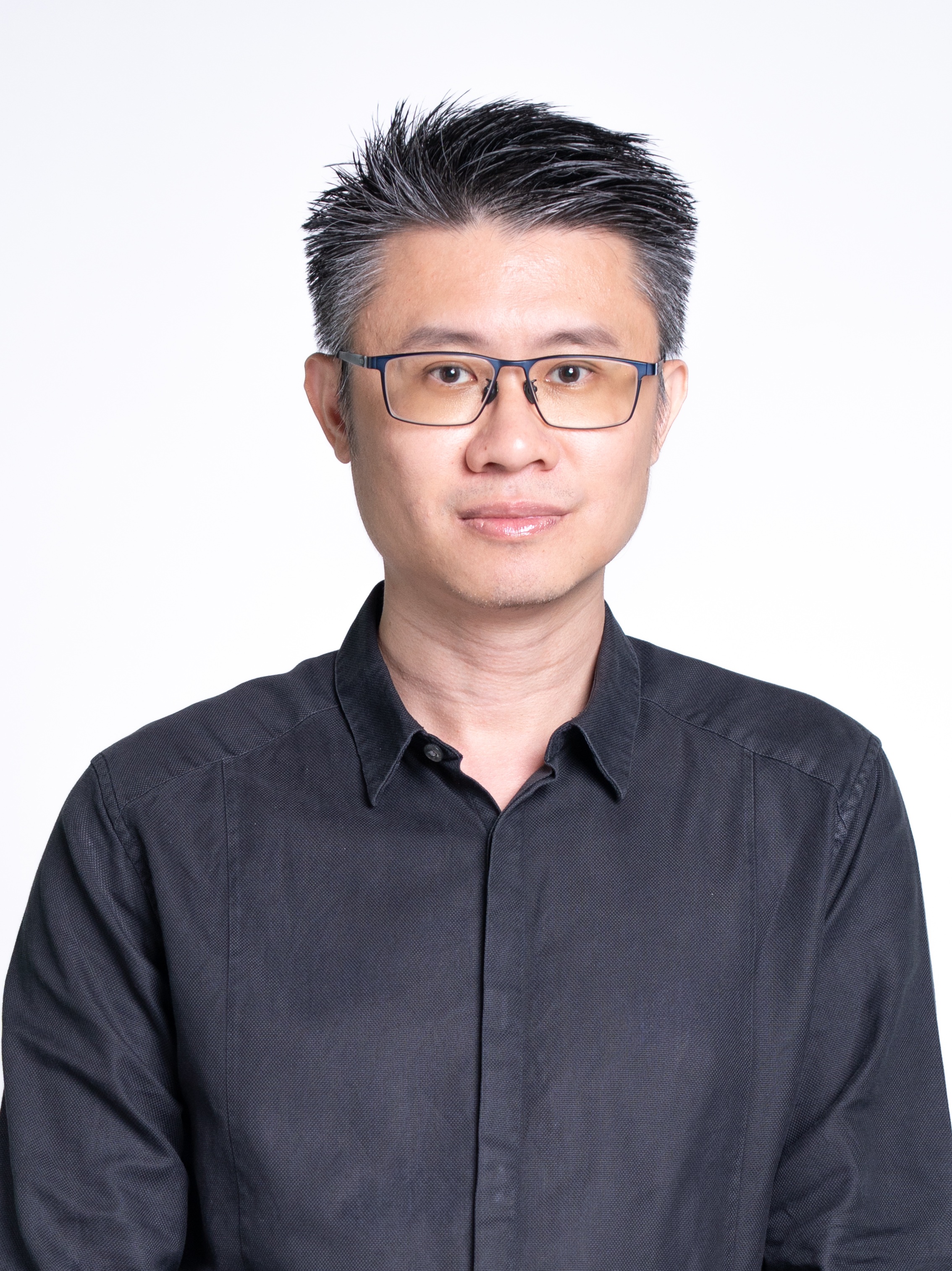}}]{Chau Yuen}
(S02-M06-SM12-F21) received the B.Eng. and Ph.D. degrees from Nanyang Technological University, Singapore, in 2000 and 2004, respectively. He was a Post-Doctoral Fellow with Lucent Technologies Bell Labs, Murray Hill, in 2005. From 2006 to 2010, he was with the Institute for Infocomm Research, Singapore. From 2010 to 2023, he was with the Engineering Product Development Pillar, Singapore University of Technology and Design. Since 2023, he has been with the School of Electrical and Electronic Engineering, Nanyang Technological University, currently he is Provost’s Chair in Wireless Communications, Assistant Dean in Graduate College, and Cluster Director for Sustainable Built Environment at ER@IN.
 
Dr. Yuen received IEEE Communications Society Leonard G. Abraham Prize (2024), IEEE Communications Society Best Tutorial Paper Award (2024), IEEE Communications Society Fred W. Ellersick Prize (2023), IEEE Marconi Prize Paper Award in Wireless Communications (2021), IEEE APB Outstanding Paper Award (2023), and EURASIP Best Paper Award for JOURNAL ON WIRELESS COMMUNICATIONS AND NETWORKING (2021).
 
Dr Yuen current serves as an Editor-in-Chief for Springer Nature Computer Science, Editor for IEEE TRANSACTIONS ON VEHICULAR TECHNOLOGY, IEEE TRANSACTIONS ON NEURAL NETWORKS AND LEARNING SYSTEMS, and IEEE TRANSACTIONS ON NETWORK SCIENCE AND ENGINEERING, where he was awarded as IEEE TNSE Excellent Editor Award 2025, 2024 and 2022, and Top Associate Editor for TVT from 2009 to 2015. He also served as the guest editor for several special issues, including IEEE JOURNAL ON SELECTED AREAS IN COMMUNICATIONS, IEEE WIRELESS COMMUNICATIONS MAGAZINE, IEEE COMMUNICATIONS MAGAZINE, IEEE VEHICULAR TECHNOLOGY MAGAZINE, IEEE TRANSACTIONS ON COGNITIVE COMMUNICATIONS AND NETWORKING, and ELSEVIER APPLIED ENERGY.
 
He is listed as Top 2\% Scientists by Stanford University, and also a Highly Cited Researcher by Clarivate Web of Science from 2022.
\end{IEEEbiography}

\end{document}